\documentclass[11pt]{elsarticle}
\usepackage{graphicx} 
\graphicspath{{figures/}}
\usepackage{amsfonts} 
\usepackage{amssymb}  
\usepackage{amsmath}
\usepackage{hyperref}
\usepackage{algorithm}
\usepackage{algorithmic}
\usepackage{amsthm}
\usepackage{color}
\usepackage{xcolor}
\usepackage{epstopdf}
\usepackage[normalem]{ulem}

\newtheorem{proposition}{Proposition}
\newtheorem{lemma}{Lemma}

\begin{document}

\begin{frontmatter}

\title{\textbf{Farthest sampling segmentation of triangulated surfaces}}

\author[vj]{Victoria Hern\'andez-Mederos}
\address[vj]{Instituto de Cibern\'etica, Matem\'atica y F\'isica, ICIMAF, La Habana, Cuba}

\author[d]{Dimas Mart\'{\i}nez-Morera}

\address[d]{Departamento de Matem\'atica, Universidade Federal do Amazonas,
       Manaus, Brazil}
\author[vj]{\\Jorge Estrada-Sarlabous}

\author[v]{Valia Guerra-Ones}
\address[v]{ Technical University of Delft, TU Delft, Delft, The Netherlands}

\begin{abstract}
In this paper we introduce \textit{Farthest Sampling Segmentation (FSS)}, a new method for segmentation of triangulated surfaces, which consists of two fundamental steps: the computation of a submatrix $W^k$ of the affinity matrix $W$ and the application of the {\it k-means} clustering algorithm to the rows of $W^k$. The submatrix  $W^k$ is obtained computing the affinity between all triangles and only a few special triangles: those which are farthest in the defined metric. This is equivalent to select a sample of columns of $W$ without
constructing it completely. The proposed method is computationally cheaper than other segmentation algorithms, since it
only calculates few columns of $W$ and it does not require the eigendecomposition of $W$ or of any submatrix of $W$.

We prove that the orthogonal projection of $W$ on the space generated by the columns of $W^k$ coincides with the orthogonal
projection of $W$ on the space generated by the $k$ eigenvectors computed by Nystr\"om's method using the columns of $W^k$
as a sample of $W$. Further, it is shown that for increasing size $k$, the proximity relationship among the rows of $W^k$
tends to faithfully reflect the proximity among the corresponding rows of $W$.

The \textit{FSS} method does not depend on parameters that must be tuned by hand and it is very flexible, since it can handle any
metric to define the distance between triangles. Numerical experiments with several metrics and a
large variety of 3D triangular meshes show that the segmentations obtained computing less than the $10\%$ of columns $W$ are
as good as those obtained from clustering the rows of the full matrix $W$.
\end{abstract}


\end{frontmatter}

\section{Introduction}

Mesh segmentation is an important ingredient of many geometric processing and computer graphics tasks, such as shape matching,
parametrization, mesh editing and  compression, texture mapping,  morphing, multiresolution modeling,  animation, etc.
It explains why this subject has received a lot of attention in recent years.
In a review of mesh segmentation techniques, Shamir \cite{Sha08} formulates the segmentation problem as an optimization problem
and considers two qualitatively different types of segmentation: the part-type,  aiming to partition the surface into volumetric
parts and the surface-type, attempting to segment the surface into patches. Segmentation techniques are also classified in
correspondence with general clustering algorithms, such as region growing, hierarchical clustering, iterative clustering,
spectral analysis, etc.

The most important tasks concerning shape segmentation is how to define a part of the surface. This is done by using various mesh
properties or features such as area, size or length, curvature,  geodesic distances, normal directions, distance to the medial axis
and shape diameter.  In many segmentation algorithms \cite{KT03}, \cite{dGGV08}, \cite{LZ04}, \cite{LLSCS05}, \cite{PKA03},
\cite{ZL05},  part analysis is carried out by using surface based computations. For instance, geodesic and angular distances are
combined in \cite{ZL05} to define a metric that encodes distances between mesh faces, while diffusion distance is used in
\cite{dGGV08} to propose a hierarchical segmentation method for articulated bodies. The previous approaches are based on
intrinsic metrics on the surfaces and do not capture explicitly volumetric information. In \cite{LZSC09} a volumetric part-aware
metric is defined, combining the volume enclosed by the surface with geodesic and angular distances.  This  metric is successfully
applied in various applications including mesh segmentation, shape registration, part-aware sampling and shape retrieval.

According to the underlying technique, many segmentation algorithms \cite{LZ04}, \cite{ZL05}, \cite{LJZ06}, \cite{LZ07},
\cite{dGGV08}, \cite{KLT05}  belong to the class of the so called spectral methods. In these algorithms, an affinity or Laplacian
matrix is constructed by using intrinsic metrics. The original surface is projected into low dimensional spaces, which are derived
from the eigenvectors of the affinity or Laplacian matrix. As a consequence of the Polarization Theorem,  higher-quality cut
boundaries may be obtained from these embeddings. For details about the spectral approach for mesh processing and analysis,
including mesh compression, correspondence, parameterization, segmentation, surface reconstruction, and remeshing, see the excellent
survey \cite{ZVKD10}.

\subsection{Contributions}

The main contribution of this paper is a new algorithm for segmentation of triangulated surfaces based on the computation of few columns of the affinity matrix $W$. Given a distance between neighboring faces, the affinity among all faces and a sample of few distinguished faces is computed. The segmentation is performed applying a classical clustering algorithm to the rows of the matrix $W^k$ composed by the sample of the few $k$ columns of $W$. The new method, called \textit{Farthest Sampling Segmentation (FSS)} does not require to compute the spectrum of $W$ or of any of its submatrices. Hence, it is computationally cheaper than the segmentation algorithms based on eigendecompositions.

From the theoretical point of view our first result is the proof that given a sample $W^k$ of the columns of  $W$, the orthogonal projection of $W$ in the space generated by the columns of $W^k$ is the same as the orthogonal projection of $W$ in the space generated by the approximated eigenvectors of $W$, obtained using Nystr\"{o}m's method for the same sample. This theoretical result clarifies the point of contact between our method and the spectral approach and explains the success of the new method \textit{FSS}. Moreover, it is shown that
if the columns of $W^k$ correspond to the $k$ farthest triangles in the selected metric, then for increasing size $k$,  pairs of faces which are close in the selected metric project in pairs of rows of order $n \times k$ matrix $W^k$ that are close as points in $\mathbb{R}^k$ and also pairs of faces which are far-away in the selected metric project in pairs of rows of $W^k$ that are far-away as points in $\mathbb{R}^k$.

A wide experimentation that illustrates the quantitative and  qualitative performance of the \textit{FSS} method is also included. Through these experiments it becomes apparent the robustness of the method for small samples of $W$. Furthermore, it is proposed  an estimate of lower bound for the size $k$ of the sample that furnishes a good approximation to $W$.

\subsection{Paper organization and notation}
In section \ref{distance} we introduce the basic concepts that allow to define the similarity between any two triangles of the mesh.
Two low dimensional embeddings of the affinity matrix are described in section \ref{embbedings}: the spectral and the statistical
leverage. The main theoretical result of the paper is included in this section. The embedding based on the computation of the columns of the affinity matrix that corresponds to the farthest triangles is introduced in section \ref{algor}, where the coherence and stability of \textit{FSS} method is shown and the associated algorithm is explained. Section \ref{NumericalExp} is devoted to the numerical experiments, which show the quantitative and qualitative performance of \textit{FSS}. Moreover, this section also includes an experiment to prove that the immersion based on the farthest
triangles provides a good approximation of $W$. A comparison of \textit{FSS} with the spectral method is finally included . The last section concludes the paper.
We use capital letters to denote matrices and the same lower case letter to denote its elements. For example the element $i,j$ of
the matrix $A$ is denoted by $a_{ij}$. Moreover, $A_{i,\cdot}$ and $A_{\cdot,j}$ represents the $i$-th row and the $j$-th column
of matrix $A$ respectively, while $A^+$ denotes the Moore-Penrose inverse of  matrix $A$.
All mesh segmentations shown in this work are computed without including any procedure to improve the smoothness of the boundaries of the segments or their concavity, such as proposed in \cite{SSCO08},  \cite{WTKL14}.


\section{Distance and affinity matrices}\label{distance}

Denoting by $T$ the triangulation composed by a set $F$ of faces $f_i,\;i=1,...,n$, the segmentation problem consists
in defining a partition of $F$. In most of the segmentation algorithms, an important step to group the elements of $F$
consists in introducing pairwise face distances and constructing an affinity matrix by using them. In the literature, several
face distances have been considered, see \cite{KT03, GSCO07, SSCO08,LZSC09}. For instance, in \cite{KT03} the distance between
two adjacent faces is defined as a convex combination of their geodesic and angular distances. Other metrics have been specially
designed to capture parts of the volume enclosed by the surface, such as the part-aware distance \cite{LZSC09} and the
shape-diameter function ({\it SDF}) \cite{SSCO08}. The part-aware metric in \cite{LZSC09} happens to be expensive, since its computation
requires to perform two samplings of the triangular mesh by using ray-shooting. On the other hand, as remarked in \cite{LZSC09},
the {\it SDF} function does not capture well the volumetric context.

\subsection{Distance matrix}

Denote by $f_i$ and $f_j$ two triangles of $T$ \textit{sharing} an edge. Assume that we have already defined a {\it distance} $d_{ij}$ between faces $f_i$ and $f_j$. For instance $d_{ij}$ could be the {\it angular distance} defined as
$\eta(1-\langle n_i,n_j \rangle)$, where $\langle n_i,n_j \rangle$ denotes the scalar product between the
normalized normal vectors $n_i,n_j$ to the triangles $f_i$ and $f_j$ respectively and $\eta$ is a weight introduced to reinforce the concavity of the angles. Another distance very common in the literature is the {\it geodesic distance}, that in the case of neighboring triangles is defined as the length of the shortest path between their barycenters $b_i$ and $b_j$, see the details on these two distances in \cite{KT03}. Our third test distance is introduced in \cite{LZSC09} and is based on  a scalar function defined on the
triangulation, the so called \textit{SDF}-function (see \cite{SSCO08}). The {\it sdf} distance between any two adjacent
faces  $f_i$ and $f_j$  of $T$ is defined as $|SDF(b_i)-SDF(b_j)|$.  Each of these test distances captures different
features of the triangulation.

The distance $d_{ij}$ between {\it any} pair of faces $f_i$ and $f_j$ is computed using the {\it weighted dual graph} $G_d$
defined from the selected distance. The $i$-th knot of the graph $G_d$ represents the triangle $f_i$ in $T$ for $i=1,...,n$,
and there is an edge between the $i$-th and the $j$-th nodes of the graph $G_d$ if the faces $f_i$ and $f_j$ share an edge on the
triangulation $T$. The weight of the edge joining the $i$-th and the $j$-th nodes in $G_d$ is $d_{ij}$ if $d_{ij} > 0$ and otherwise it is set equal to $\varepsilon$, with $0 < \varepsilon$ very small. The distance $d_{ij}$ between faces $f_i$ and $f_j$, which are not necessarily neighboring faces, is defined as {\it the length of the shortest path} between the $i$-th and the $j$-th knots in the graph $G_d$. This length may be computed using Dijkstra algorithm. Observe that the distance $d_{ij}$ satisfies the axioms of metric.
We denote by $D=(d_{ij}),\;i,j=1,...,n$ the matrix of the distances between each pair of faces of the triangulation.

\subsection{Affinity matrix}\label{sectionAm}

Given a suitable metric that allows to compute the pair-wise distance between faces of the triangulation, the affinity matrix $W$
encodes the probability of each pair of faces of being part of the same cluster and can be considered as the adjacency matrix
of the weighted graph $G_d$ previously introduced.

Assume that the distance $d_{ij}$ between any pair of faces $f_i$ and $f_j$ of the triangulated surface has been already
computed. Then the affinity $w_{ij}$ between  faces $f_i$ and $f_j$  which are closer should be large. In the literature it
is customary to use a {\it Gaussian kernel} to define  $w_{ij},\;i,j=1,...,n$  as
\begin{equation}
w_{ij}=e^{-d_{ij}/(2\sigma^2)}
\label{afinidad}
\end{equation}
where $\sigma=\frac{1}{n^2}\sum_{i}\sum_{j} d_{ij}$. Observe that $0<w_{ij} \leq 1$ and  $w_{ii}=1$ for all $i=1,...,n$. Moreover,
$W=(w_{ij}),\;i,j=1,...,n$ is a symmetric matrix. Denote by $M$ the diagonal matrix $M=diag(m_{ii})$, where $m_{ii}=\sum_{j=1}^n w_{ij}$.

Many papers in the literature deal with normalized versions of affinity matrix, which  are called Laplacians  in the more
general context of clustering of data for exploratory analysis, see \cite{vL06}. For instance,   in \cite{SM97} the (nonsymmetric)
affinity matrix $M^{-1}W$ is used in spectral image segmentation, while the symmetric normalized affinity matrix $Q=M^{-1/2}WM^{-1/2}$ is used in \cite{NgW02} for data clustering and  in \cite{LZ04} to segment triangular meshes. In applications, the affinity matrix $W$ of order $n$ is huge, therefore segmentation methods requiring the computation of all matrix entries are very expensive. To overcome this problem,
the segmentation algorithm proposed in this paper computes only few columns of  matrix $W$.


\section{Low dimensional embeddings for clustering }\label{embbedings}

In geometric processing community, low dimensional embeddings are frequently used to transform the input data from its
original domain to another domain. The main purpose of these embeddings is to reduce the dimensionality of the problem,
preserving the information of the original data in such a way that the solution of the new problem is cheaper and easier.
The segmentation problem can also be considered as a clustering problem, which is frequently solved applying
{\it k-means} method, \cite{Lloyd82}. In this context, dimensionality reduction for {\it k-means} is strongly connected with
low rank approximation of the matrix containing the data to be clustered \cite{BZMD15}. In our problem, the matrix containing
the information about ``data points'' is the affinity matrix $W$. Each row of $W$ represents the affinity between a triangular
face and the rest of faces. Hence, a valid strategy to solve the segmentation problem consists in computing a low rank
approximation of the affinity matrix $W$ and clustering its rows.

\subsection{Spectral approach}\label{SS}
The most popular low dimensional embeddings in the literature are the spectral ones, which are constructed from a set of eigenvectors of a
properly defined linear operator. They have been successfully applied in mesh segmentation \cite{LZ04},\cite{LJZ06}, \cite{LZ07},
\cite{dGGV08} and also in other geometric processing applications, such as shape correspondence \cite{JZK07} and retrieval \cite{EK03}
and mesh parametrization \cite{Got03}, \cite{MTAD08}.

From the theoretical point of view, spectral embeddings are supported by a classical linear algebra result, the Eckart-Young
theorem \cite{EY36}. It establishes that the best rank $k$ approximation in the Frobenius norm of a real, symmetric and
positive semi-definite matrix $W$ of dimension $n$ is the matrix
\begin{equation}
E^k=\widetilde{U}^k(\widetilde{U}^k)^t
\label{eigmatrix}
\end{equation}
where $\widetilde{U}^k$ is the matrix with columns $\sqrt{\lambda_1}u_1,\sqrt{\lambda_2}u_2,...,\sqrt{\lambda_k}u_k$ and
$u_1,u_2,...,u_k$ are the eigenvectors of $W$ corresponding to its largest eigenvalues $\lambda_1 \geq \lambda_2 \geq ...\geq \lambda_k$.
It means that the following equality holds for the Frobenius norm of the error $W-E^k$
\begin{equation}
\|W-E^k\|_F =\min_{X\in \mathbb{R}^{n \times k} ,\;rank(X) \leq k}\|W-XX^{t}\|_F
\label{mejoraprox}
\end{equation}

If we denote by $U^k$ the matrix with columns  $u_1,u_2,...,u_k$, then $\widetilde{U}^k=(\Lambda^k)^{\frac{1}{2}}U^k$,
where $\Lambda^k$ is the order $k$ diagonal matrix with diagonal elements $\lambda_1,\lambda_2,...,\lambda_k$. Moreover, it not difficult
to prove that $(U^k)^{+}=(U^k)^{t}$. Hence, the orthogonal projection $U^{k}(U^k)^{+}W$ of $W$ on the space generated by columns of $U^k$ satisfies
\begin{eqnarray*}
U^{k}(U^k)^{+}W&=&U^{k}(U^k)^{t}W=U^{k}(\Lambda^k)(U^k)^{t}\\
               &=&(\widetilde{U}^k(\Lambda^k)^{-\frac{1}{2}})(\Lambda^k)(\widetilde{U}^k(\Lambda^k)^{-\frac{1}{2}})^{t}\\
               &=&\widetilde{U}^k(\widetilde{U}^k)^{t}=E^k
\end{eqnarray*}
The equality $E^k=U^{k}(U^k)^{+}W$ means that the best rank $k$ approximation of $W$ is the projection of $W$ on the space
generated by the eigenvectors of $W$ corresponding to its largest eigenvalues.

Spectral clustering algorithms also rely on the polarization theorem \cite{BH03} which suggests that as the
dimensionality of the  spectral embeddings decreases the clusters in the data are better defined. In practical applications,
it is necessary to choose a value of $k$ representing a good compromise between these two apparently conflicting results. This value
should be small enough to obtain a good polarization of the embedding data, but at the same time large enough to reduce the
distortion of the relationships among the data due to the embedding.

Spectral methods of segmentation are in general expensive, since they require the computation of eigenvalues and eigenvectors
of the  so called Laplacian matrix.  In some cases \cite{LZ04},\cite{LJZ06}, the Laplacian matrix is obtained introducing a normalization
of the affinity matrix. In other cases \cite{RBGPS09}, it arises from a discretization of the Laplace-Beltrami operator. In geometry processing context
the Laplacian matrix used in segmentation  is a dense and usually very large matrix. To face this problem, \cite{LJZ06} uses
Nystr\"{o}m's method, since it only requires a small number of sampled rows of the affinity matrix and the solution of a small scale
eigenvalue problem. More precisely, the set of faces of $F$ is partitioned in two: a set $\mathcal{X}$ of the faces of the ``sample"
of size $k << n$ and its complement $\mathcal{Y}$ of  size $n-k$. Let be $p_{\mathcal{X}}$ the set of $k$ indices of the faces
contained in the sample $\mathcal{X}$, $p_{\mathcal{Y}}$
the set of $n-k$ indices of the faces contained in  $\mathcal{Y}$ and $p=( p_{\mathcal{X}}, p_{\mathcal{Y}})$ the vector
representation of the permutation matrix $P$, then the permuted affinity matrix $W_P:=PWP^t$ has the following structure
\begin{equation}
W_P:=PWP^t=\left[
    \begin{array}{cc}
      A & B \\
      B^{t} & C\\
    \end{array}
  \right]
  \label{Wp}
\end{equation}

 where $A$ is the order $k$ affinity matrix of the elements in $\mathcal{X}$ and  $B$ is the order $k \times n$ matrix of
 the cross-affinities between elements in $\mathcal{X}$ and  $\mathcal{Y}$. The eigenvectors of $W$ corresponding to the $k$
 largest eigenvalues, i.e. the columns of $U^k$, may be approximated  \cite {FBCM04}, \cite{LJZ06} by the columns of the $n \times k$  matrix $P^t \, N^k$, with
 \begin{equation}
N^k=\left[
    \begin{array}{c}
      U_A \\
      B^{t}U_A\Lambda_A^{-1}\\
    \end{array}
  \right]
\label{Nk}
\end{equation}

where $A=U_A \Lambda_A U_A^{t}$ is the spectral decomposition of $A$. The orthogonal projection  $F^k$  of $W$ on the space generated by the columns of matrix $P^tN^k$ provided by Nystr\"{o}m's method is given by
\begin{equation}
F^k= (P^tN^k)\,(P^tN^k)^{+}W
\label{proyNystrom}
\end{equation}

The accuracy of the eigenvectors computed by using the Nystr\"{o}m's method strongly depends on the
selection of the $k$ columns of $W$ corresponding to sample $\mathcal{X}$. Therefore, different schemes have been considered in the literature,
for instance random sampling, uniform sampling, max-min farthest point sampling  and greedy sampling \cite{FBCM04}, \cite{KMA12},
\cite{Mah11},\cite{ST04},\cite{LJZ06}.\\

Nystr\"{o}m's method has associated an approximation to $W_P$ given by
\begin{equation}
\widetilde{W}_P:=N^k\Lambda_A(N^k)^t
\label{Wponda}
\end{equation}

It is straightforward \cite{FBCM04} to check that it holds
\begin{equation}
\widetilde{W}_P=\left[
    \begin{array}{cc}
      A & B \\
      B^{t} & B^{t}A^{-1}B\\
    \end{array}
  \right].
\label{Wponda1}
\end{equation}

Since $\widetilde{W}_P \approx W_P$, after  \ref{Wp}, \ref{proyNystrom} and \ref{Wponda} we have that
\begin{eqnarray*}
F^k & = & P^tN^k(N^k)^+PW(P^tP)= P^tN^k(N^k)^+W_PP\\
 &\approx &  P^tN^k(N^k)^+\widetilde{W}_PP\\
 &=&  P^tN^k(N^k)^+(N^k\Lambda_A(N^k)^t)P\\
 &=& P^tN^k\Lambda_A(N^k)^tP= P^t\widetilde{W}_PP \\
 &\approx& P^tW_PP=  P^t(PWP^t)P =W
\end{eqnarray*}

Hence, $F^k$ may be considered as an approximation to the affinity matrix $W$ with similar quality  as the Nystr\"{o}m approximation  $ P^t\widetilde{W}_PP$ to $W$. The accuracy of the approximation $F^k$ also depends on the
selection of the $k$ columns of $W$ corresponding to sample $\mathcal{X}$.\\

\begin{proposition}
Let $W$ be a symmetric order $n$ matrix and  $p=( p_{\mathcal{X}},p_{\mathcal{Y}} )$ a permutation vector of indices 1,2,...,n,
where $ p_{\mathcal{X}}$ has size $k$. Let be $P$ the order $n$ permutation matrix represented by vector $p$ and
$W^k$ the $n \times k$ matrix whose columns are the columns of $W$ with indices in $p_{\mathcal{X}}$. Then, it holds,
\begin{equation}
(W^k)\, (W^k)^{+} =(P^tN^k)\,(P^tN^k)^{+}
\label{igualproy}
\end{equation}
where $N^k$ is given by (\ref{Nk}).
\end{proposition}

\textbf{Proof}. \\
From $A=U_A \Lambda_A U_A^t$ it follows $U_A=A U_A \Lambda_A^{-1}$. Hence,
 \begin{equation}
N^k=\left[
    \begin{array}{c}
      A U_A \Lambda_A^{-1} \\
      B^{t}U_A\Lambda_A^{-1}\\
    \end{array}
  \right] =\left[
    \begin{array}{c}
      A \\
      B^{t}\\
    \end{array}
  \right]U_A\Lambda_A^{-1}\\
\end{equation}
But $\left[
    \begin{array}{c}
      A \\
      B^{t}\\
    \end{array}
  \right]=P W^k $,
  thus it holds $N^k=P W^k U_A \Lambda_A^{-1}$ and we get $$(P^t N^k)(P^tN^k)^{+}=(W^k U_A \Lambda_A^{-1})(W^k U_A \Lambda_A^{-1})^{+}= (W^k)(W^k)^{+}.  \hspace{3cm} \blacksquare $$
\smallskip
{\bf Remarks}\\
From the previous result it holds that:
\begin{itemize}
\item  The orthogonal projection $H^k$ of $W$ on the space generated by the columns of $W^k$  given by
\begin{equation}
H^k = W^k{(W^k) }^+ W
\label{Hk}
\end{equation}
coincides with  the orthogonal projection $F^k$ of $W$ on the space generated by the columns of matrix $P^tN^k$, which is provided by Nystr\"{o}m's method.
\item Since $H^k=F^k$, the accuracy of the approximation $H^k$ to $W$ is similar to Nystr\"{o}m approximation $ P^t\widetilde{W}_PP$ to $W$. It also depends on the selection of the $k$ columns of $W$ corresponding to sample $\mathcal{X}$.
\item If we associate the $i$th-face of $T$ with the $i$th-row of $W$,
then clustering the rows of $W$ may be replaced by clustering either the rows of $W^k$ or the rows of $P^t N^k$.
\end{itemize}

\subsection{Statistical ``leverage" approach}

Mahoney proposes in \cite{Mah11} a method to select a ``sample'' of the columns of a matrix $W$ of dimension $n$ in such
a way that the space generated by the selected columns provides a good approximation of $W$. Given $k$, with $k <<n $,
the method assigns to the $j$-th column of $W$ a ``leverage'' or ``importance score'' $\pi_j$ that measures  the influence of
that column in the best rank $k$ approximation of $W$.

The use of the leverage scores for column subset selection dates back to 1972, \cite{Jol72}. However, the introduction of the
randomized approach has given an essential theoretical support, \cite{HIW15}, to the leverage scores in the role of revealing
the important information hidden in the underlying matrix structure.

More precisely, if $v_1,v_2,...,v_k$ are the right singular vectors of $W$ corresponding to the largest singular values, the leverage
$\pi_j$ is defined as
\begin{equation}
\pi_j=\frac{1}{k}\sum_{i=1}^k (v_i^j)^2
\label{leveragej}
\end{equation}

where $v_i^j$ denotes the $j$-th component of $v_i$. The normalization factor $\frac{1}{k}$ is introduced in (\ref{leveragej})
to consider $\pi_j$ as a probability associated to the $j$-th column of $W$.  By using that score as an importance sampling
probability distribution, the algorithm in \cite{Mah11}  constructs a $n \times m$ matrix $C$, composed by $m \geq k$
columns of $W$. With high probability, the selected columns are those that exert a large influence
on the best rank $k$ approximation of $W$.
In our experiments in section \ref{NumericalExp}, we use a slight modification of Mahoney's algorithm to obtain a matrix $C$,
that we denote by $C^k$, that is composed exactly by $k$ columns. More precisely, if we arrange in decreasing order the
leverages  $\pi_{j_1} \geq \pi_{j_2} \geq ... \geq \pi_{j_n} $
then, the $i-$th column of matrix $C^k$ is the column $j_i$ of $W$, for $i=1,...,k$.
The orthogonal projection of $W$ on the space generated by the columns of $C^k$ is given by,

\begin{equation}
G^k=C^k\,(C^k)^{+}W
\label{proyleverage}
\end{equation}

\section{Farthest sampling mesh segmentation method (\textit{FSS})} \label{algor}
Computing distances from every node to a subset of nodes of a graph (landmarks or reference objects) is a well known method to efficiently provide estimates of the actual distance. In this context, this distance information is also referred to as an embedding. Landmarks have been used for graph measurements in many applications, such as roundtrip propagation, transmission delay or social search
in networks, but the optimal selection of the location of landmarks has not been exhaustively studied [PBCG09]. Kleinberg et al.
\cite{KSW04} address the problem of approximating network distances in real networks via embeddings using a small set of  landmarks and obtain theoretical bounds for  choosing landmarks randomly. More recently, in [PBCG09] is shown that  in practice, simple intuitive strategies outperform the random strategy.  \\

The previous ideas and the result and discussion at the end of section \ref{SS} inspired us to propose a mesh segmentation  method based
on the computation of a small sample of columns of the affinity matrix $W$. It is well known  that the quality of the approximation $P^t\widetilde{W}_PP$ to $W$ given by Nystr\"{o}m's method (see (\ref{Wponda1})), strongly depends on the selection of the sample. Consequently, our segmentation method {\it FSS} consists in two steps: first we  use the {\it farthest point heuristic} to select which columns of $W$ are computed and then, the {\it clustering method} is applied to the rows of this submatrix of $W$.\\

The rationality behind these steps is the following. A representative subset of the columns of $W$ must have maximal rank. Since the
entries of the $j$-th column  of $W$ are the affinity values between all faces of $T$ and the $j$-th face, the sample should not include columns corresponding to faces which are very close between them.  In this sense, the \textit{ farthest point heuristic} is a good strategy to avoid redundancy in the sample. On the other hand, if two faces $f_i$ and $f_j$ of $T$ are close in the selected metric $d$, i.e., $d_{i,j}$ is small, then the corresponding row vectors $D_{i,\cdot}$ and $D_{j,\cdot}$ of the whole distance matrix $D$ are approximately equal, since according to the triangular inequality the difference between the $r$-th  components of $D_{i,\cdot}$ and $D_{j,\cdot}$, is bounded by $d_{i,j}$ for $r=1,...,n$. Hence, due the continuity of the Gaussian kernel, the row vectors $W_{i,\cdot}$ and $W_{j,\cdot}$ of the affinity matrix $W$ are also approximately equal with a difference greater or equal than the difference between the $i$-th and $j$-th rows of $W^k$. In other words, the proximity among rows of $W$ and consequently between the faces of $T$, is well reflected by the proximity of the same rows of $W^k$.

\subsection{Sampling procedure}
Our algorithm {\it FSS} is deterministic and greedy in the sense that at each iterative step, it makes a decision about which column to add according to a rule that depends on the  already selected columns. As we already mentioned, the sample of $W$ is derived from a sample
of columns of the distance matrix $D$.
To obtain a good approximation of $D$ it is enough to select a set $\mathcal{X} \subset F$ of distinguished faces
that can be considered as landmarks, in the sense that the distance between any pair of faces $f_i$ and $f_j$ can be approximated
in terms of the distance of $f_i$ (respectively $f_j$) to the landmark faces.

The method iteratively computes the columns of a matrix $X$, which contains a sample of $k$ columns of the distance matrix $D$. More precisely, in the first step we choose randomly a value $j_1$ with $1 \leq j_1 \leq n$ and define the first column of matrix $X$ as the vector built up with the distances of all faces to the $j_1$-th face, $X_{i,1}={d}_{i,j_1},\,i=1,...,n$. Then, we search the index $j_2$
of the face farthest to the $j_1$-th face and assign to the second column of $X$ the vector of the distances of all faces to the
face $j_2$. In general, in the step $l,\; k \geq l \geq 2$, we have the $l-1$ indices  $j_1,j_2,...,j_{l-1}$ previously selected
and a matrix $X$ of order $n \times (l-1)$ with columns ${D}_{\cdot,j_1},{D}_{\cdot,j_2},...,{D}_{\cdot,j_{l-1}}$,  which contains the
distances of all faces $f_i,\;i=1,...,n$ to the faces $f_{j_1},f_{j_2},...,f_{j_{l-1}}$. In this step we look for the index
$j_l$ of the face that maximizes the minimal distance to the faces $f_{j_1},f_{j_2},...,f_{j_{l-1}}$,
\begin{equation}
j_l=arg \{\max_{1 \le i \le n} \{ \min_{1 \leq r \leq l-1}\; x_{i,r}\}\}
\label{jk}
\end{equation}
where $x_{i,r}= {d}_{i,j_r},\;\;i=1,...,n,\;r=1,...,l-1$ is the element of $X$ in the position $(i,r)$, i.e, the
distance between faces $f_i$ and $f_{j_r}$. Once we have $j_l$, we compute the $l$-th column  of $X$ as the vector of
distances of all faces to the $j_l$-th face.\\

The $i$-th row of $X$ are the coordinates of a point in $\mathbb{R}^k$, which could be considered  as a $k$ dimensional
embedding of the point in $\mathbb{R}^n$ given by the $i$-th row of $D$ (which represents the distances of all faces to the face $f_i$).
Given $k$, by using the matrix $X$ we compute a matrix $W^k=(w^k_{ij})$ of order $n \times k$, which is composed by a sample of
columns of the affinity matrix $W$,

\begin{equation}
w^k_{ij}=e^{-x_{ij}/(2\sigma_k^2)},\,i=1,...,n,\;\;j=1,...,k
\label{afinidad2}
\end{equation}
where
\begin{equation}
\sigma_k=\frac{1}{nk}\sum_{i=1}^n\sum_{j=1}^k x_{i,j}
\label{sigmank}
\end{equation}

Finally, it remains to explain how we compute the size $k$ of the sample. In this sense, several options are possible.
The simplest one is to define a priori the value of $k$, for instance as the integer part of prescribed percent of the total $n$
of faces. In this case, the {\it sampling algorithm} may be summarized as follows.

\medskip

\begin{algorithmic}[1]
\STATE {\bf Procedure Sampling}
\STATE {\bf input}: triangulation $T$, number of clusters $n_c$, size of sample $k$.
\STATE Choose randomly an index  $j_1$ with $1 \leq j_1 \leq n$.
\FOR{$l=1$ to $k$}
 \STATE Compute the distance of all $n$ faces to the face $j_l$. Assign the distances to the column $l$ of matrix $X$.
 \STATE Compute $j_{l+1}$ by using (\ref{jk}).
\ENDFOR
\STATE {\bf output}: sampling distance matrix $X$.
\end{algorithmic}

\medskip

Another option for computing the size $k$ of the sample is the following. A value $\beta_l \geq 0, \;l \geq 1$ strongly related to the selection of index $j_l$ in (\ref{jk}) is introduced defining,
\begin{equation}
\beta_l:=\max_{1 \le i \le n} \{ \min_{1 \leq r \leq l}\; x_{i,r}\}
\label{betak}
\end{equation}
Recall that for all $l \geq 1$, $\beta_l \geq 0$. Furthermore, from the definition  (\ref{betak}) it is clear that
the sequence  $\{ \beta_l,\;1 \le l \le n\}$ is monotonic decreasing with $\beta_n=0$. In fact, while new faces are included
into the sample, the distance of the new face that is simultaneously farthest away to all actual members of the sample decreases.
If the sample contains  all faces, i.e. if $l=n$, then matrix $X$ is a permutation of columns of the distance matrix $D$ and
for all $i=1,...,n$ it holds  $\min_{1 \leq r \leq n}\; x_{i,r}=0$, thus we get $\beta_n=0$. Hence, the value of $\beta_k$ may
be also interpreted as a measure of the error introduced when the ``original'' data in a $n$-dimensional space are substituted
by their $k$-dimensional embedding. Given an upper bound $\epsilon >0$, the size $k$ of the sample may be computed as
\begin{equation}
k=\min_{2 \le l \le n} \{ l \mbox{  such that  }  \frac{\beta_l}{\beta_1} < \epsilon\}
\label{selectk}
\end{equation}
For  $1 > \epsilon >0$, the value of $k$ computed by using (\ref{selectk}), depends on $\epsilon$ and it
is usually much smaller than $n$. A pseudocode of the previous {\it Sampling algorithm} is included below.

\medskip

\begin{algorithmic}[1]
\STATE {\bf Procedure Sampling}
\STATE {\bf input}: triangulation $T$, number of clusters $n_c$, $\epsilon>0$.
\STATE Choose randomly an index  $j_1$ with $1 \leq j_1 \leq n$, assign $l \leftarrow 1$ and $\delta \leftarrow 2\epsilon$.
\WHILE {$\delta > \epsilon$ and $l < n $}
   \STATE Compute the distance of all $n$ faces to the face $j_l$. Assign the distances to the column $l$ of matrix $X$.
   \STATE Compute $\beta_{l}$ and $j_{l+1}$ by using (\ref{betak}) and (\ref{jk}) respectively.
   \STATE Assign $\delta\leftarrow \frac{\beta_l}{\beta_1}$
   \STATE Assign $l\leftarrow l+1$.
\ENDWHILE
\STATE {\bf output}: sampling distance matrix $X$.
\end{algorithmic}

\subsection{Coherence and stability of FSS}\label{coherence}

Now we are ready to explain why \textit{FSS} is {\it coherent} in the sense that clustering the rows of $W^k$ happens to be consistent with clustering the rows of the full matrix $W$ and finally segmenting the triangulated surface. That is very important, since it is well known that in general points far-away  may project in very close points. This is not the case in the \textit{FSS} embbeding and consequently, no artifacts appear in the segmentation process. \\

From now on we call \textit{ farthest point (FP) sample} of size $k$ to the set of indices $\mathcal{X_F}^k:=\{j_1,j_2,...,j_k\} $ such that $j_1$ is randomly chosen and for $2 \leq l \leq k$, $j_l$ is the index of the face maximizing the distance to faces $j_1,j_2,...,j_{l-1} $. The \textit{FP sample} $\mathcal{X_F}^k $ is associated to the $n \times k$ matrix $X$, submatrix of $D$, and the $n \times k$ matrix $W^k$, whose columns are (approximately) the columns $j_1,j_2,...,j_{k} $ of the full matrix $W$.

\begin{lemma}\label{polarisation}
Given a triangulation $T$ and a selected metric $d$, for fixed initial face index $j_1$, let $\mathcal{X_F}^k$ be the \textit{FP} sample of size $k$. It exists $k^*$ such that, if two faces of $T$ are far-away (very close, respectively) in the metric $d$, then for all $k>k^*$, the corresponding rows of matrix $W^k$ are also far-away (very close, respectively) as points in $\mathbb{R}^k$.
\end{lemma}

\textbf{Proof}. \\
First we prove that for {\it any} sample $\mathcal{X}$, it holds that pairs of faces which are close in the selected metric $d$ project in pairs of rows of order $n \times k$ matrix $W^k$, associated to $\mathcal{X}$,  that are close as points in $\mathbb{R}^k$. Indeed, if two faces $f_i$ and $f_j$ of $T$ are close in the selected metric $d$, i.e., if $d_{i,j}$ is small, then the corresponding row vectors $D_{i,.}$ and $D_{j,.}$ of the full distance matrix $D$ are approximately equal, since according to the triangular inequality the difference between the $r$-th  components of $D_{i,.}$ and $D_{j,.}$, is bounded above by $d_{i,j}$, i.e.,
  $$ | d_{i,r}-d_{j,r} | \leq d_{i,j} \;\;\;\;\; \mbox{for} \;\; r=1,...,n.$$
Obviously, for any sample $\mathcal{X}\subset \{1,...,n\}  $  holds the same upper bound for the corresponding row vectors $X_{i,.}$ and $X_{j,.}$ of the order $n \times k$ matrix $X$. Hence, pairs of faces which are close in the selected metric $d$ project in pairs of rows of order $n \times k$ matrix $W^k$ that are close as points in $\mathbb{R}^k$, independently of the choice of sample $\mathcal{X} $.\\

On the other hand, assume that faces $f_i$ and $f_j$ of $T$ are far-away in the selected metric $d$, i.e. $d_{i,j}$ is large. Then for any $\varepsilon$ with $d_{i,j} \gg \varepsilon >0$, there exits $k^*$ such that $\beta_k < \frac{\varepsilon}{2}$ for all $k> k^*$. Moreover,  it is clear from the definition of $\beta_k$ that given face $f_i$, we can find an index $i^* \in \mathcal{X_F}^k$ such that 
$d_{i,i^*} \leq d_{j,i^*} \leq \beta_k$ for $1 \leq j \leq n$.
Hence,
 $$d_{j,i^*}+\beta_k \geq d_{j,i^*}+ d_{i,i^*} \geq d_{j,i}$$
From the previous inequalities we obtain that for $r  \in \mathcal{X_F}^k$, the maximum difference between the $r$-th components of $D_{i,.}$ and $D_{j,.}$, is bounded below by

$$\max_{r  \in \mathcal{X_F}^k} \{\,| d_{j,r}-d_{i,r} | \,\} \geq d_{j,i^*}- d_{i,i^*} \geq  d_{j,i}-\beta_k - d_{i,i^*} \geq d_{j,i} - 2 \beta_k > d_{j,i} - \varepsilon$$

Thus, for $k>k^*$, rows $i$ and $j$ of matrix $W^k$ are far-away, i.e., pairs of faces which are far-away  in the selected metric $d$ project in pairs of rows of matrix $W^k$ associated to \textit{FP} sample $\mathcal{X_F}^k $, that are far-away as points in $\mathbb{R}^k$. $\blacksquare$ \\

In the step 3 of $Sampling$ $Algorithm$ the initial face $j_1$ is chosen at random. Now we show that choosing at random different indices of the initial triangle, the differences between the $i$-th and $l$-th rows of the corresponding affinity matrices associated to the \textit{FP}  samples  tend to be equal for increasing size $k$. Thus, the segmentation result is {\it stable} with respect to the selection of the initial face $j_1$.

\begin{lemma}
Given a triangulation $T$ and a selected metric $d$, let be $\mathcal{X_F}^k$ and $\overline{\mathcal{X}}_{\mathcal{F}}^{\,k}$ two \textit{FP} samples of size $k$, with different initial faces $j_1$ and $\overline{j}_1$ and associated affinity matrices $W^k$ and $\overline{W}^k$, respectively. Then for increasing size $k$ it holds
$$ \|W^k_{l,.}-W^k_{i,.}\| - \| \overline{W}^k_{l,.}-\overline{W}^k_{i,.}  \| \rightarrow 0$$
for any pair of indices $l,i \in \{1,2,...,n\}$.
\end{lemma}
\textbf{Proof}. \\
Denote by $\beta_k$ and $\overline{\beta}_k$ the $\beta$ values (\ref{betak}) corresponding to $\mathcal{X_F}^k=\{j_1,...,j_k\}$ and $\overline{\mathcal{X}}_{\mathcal{F}}^{\,k}=\{\overline{j}_1,...,\overline{j}_k\}$, respectively. Set $\widetilde{\beta}_k= \max \, \{ \beta_k, \overline{\beta}_k \}$.  Given $\varepsilon >0$, if $k$ is sufficiently large,  we may assume that $\widetilde{\beta}_k < \overline{}\frac{\varepsilon}{2}$. Then, for $1 \leq r \leq k$ it holds $d_{i,j_r} \leq \widetilde{\beta}_k$ and
$d_{l,\overline{j}_r} \leq \widetilde{\beta}_k$ with $1 \leq l,i \leq n$. Hence,

$$|d_{i,j_r}-d_{i,\overline{j}_r}| \leq \widetilde{\beta}_k \hspace{2cm} \mbox{and}
|d_{l,j_r}-d_{l,\overline{j}_r}| \leq \widetilde{\beta}_k $$
 
 Thus, from the previous inequalities it follows  for $ 1 \leq l,i \leq n$ and $ 1 \leq r \leq k$,
$$|\,|d_{i,j_r}-d_{l,j_r}| - |d_{i,\overline{j}_r}-d_{l,\overline{j}_r}| \,| \leq |d_{i,j_r}-d_{i,\overline{j}_r} - d_{l,j_r}+ d_{l,\overline{j}_r} | \leq |d_{i,j_r}-d_{i,\overline{j}_r}| + |d_{l,j_r}-d_{l,\overline{j}_r}|\leq 2\widetilde{\beta}_k < \varepsilon $$

Hence, even choosing at random different indices of the initial triangle, for increasing size $k$ of the corresponding \textit{FP} samples, the difference between the $l$-th and $i$-th rows of affinity matrix $W^k$ and the differences between the $l$-th and $i$-th rows of affinity matrix $\overline{W}^k$ tend to be equal. $\blacksquare$

In section \label{section5.4} 5.4 we show that choosing fixed initial triangle $j_1$, most of the segmentation  results obtained with \textit{FP} samples of relatively small size $k$  are very close to the segmentation obtained from the full affinity matrix.\\

\subsection{Pseudocode of FSS method}

As previously pointed out, the {\it FSS} method firstly computes a small sample $W^k$ of columns of $W$. Based on the identification
of the rows of $W$ with the faces of $T$, the segmentation of the mesh is obtained clustering the rows of $W^k$. Below we include a pseudocode of the algorithm {\it FSS}, which receives as input the triangulation $T$ and the number $n_c$ of desired clusters.
The number $k$ of columns of the affinity matrix to be computed depends on the sampling procedure. In any case, it is assumed that
$k << n$. The algorithm {\it FSS} calls to the procedure \textbf{Sampling}, which returns the matrix $X$. The $l$-th column of $X$
is the vector of distances of all faces to the $j_l$-th face in the selected sample.\\

\medskip

\begin{algorithmic}[1]
\STATE {\bf Procedure FSS}
\STATE {\bf input}: triangulation $T$,number of clusters $n_c$
\STATE Call to the procedure \textbf{Sampling} to compute the sampling distance matrix $X$.
\STATE Assign $k$ as the number of columns of $X$.
\STATE Compute $j_{l+1}$ by using (\ref{jk}).
\STATE Compute the normalization factor $\sigma_k$ given by (\ref{sigmank}).
\FOR{$i=1$ to $n$}
  \FOR{$j=1$ to $k$}
     \STATE $w^k_{ij}=e^{-x_{ij}/(2\sigma_k^2)}$
  \ENDFOR
  \STATE Compute $normW_i:=\|W^k_{i,\cdot}\|_2$, where $W^k=(w^k_{ij}),\;i=1,...,n,\;j=1,...,k$
  \FOR{$j=1$ to $k$}
     \STATE  $w^k_{ij}=w^k_{ij}/normW_i$
   \ENDFOR
\ENDFOR
\STATE Apply {\it k-means} to the $n$ points given by rows of $W^k$ to obtain $n_c$ clusters.
\STATE Construct the segmentation vector $s=(s_1,...,s_n)$, where $s_i$ with $1 \leq s_i \leq n_c$, is the index of the cluster
assigned to $i$-th face of $T$, which is defined as the cluster assigned by {\it k-means} to $i$-th row of $W^k$.
\STATE {\bf output}: Segmentation vector $s=(s_1,...,s_n)$.
\label{meshsegmentation}
\end{algorithmic}

\vspace{0.5cm}
Observe that algorithm {\it FSS} normalizes the rows of $W^k$ (as suggested in \cite{LZ04}), hence these rows may be interpreted as points on the $(k-1)$-dimensional sphere $\mathbb{S}^{k-1}$. The segmentation is carried out applying the classic {\it k-means} clustering algorithm to these points. Compared with other algorithms reported in the literature, the new algorithm has several advantages. First, like \cite{LJZ06} only a submatrix $W^k$ of $W$ has to be computed. On the other hand,
unlike as in  \cite{LZ04} and \cite{LJZ06}, algorithm {\it FSS} does not require to compute the spectrum of $W$
or the spectrum of any submatrix of $W$.

The construction of matrix $W^k$ requires the computation of
$$(n-1)+(n-2)+...+(n-k)=1/2\;k(2n-k-1)=O(kn)$$
distances between faces. In comparison,  obtaining  the whole  affinity matrix $W$ is much more expensive and would require
the computation of 
$O(n^2)$ distances between faces.

The total computational cost $C_t$ of algorithm {\it FSS} is the sum of the cost $C_k$
of computing the approximation $W^k$ of the affinity matrix $W$ plus the cost  $C_s$ of clustering the rows of $W^k$ in $n_c$ clusters.
If $m$ is the cost of computing the distance $d$ between two faces of a given triangulation with $n$ faces (this cost strongly
depends of the underlying metric, for instance if the metric is the geodesic or the angular metric, then $m=O(n \log(n))$ ), then
$C_k=O(knm)$. Since $k$, $n_c$ and the number $n_i$ of Lloyd  \cite{Lloyd82} iterations  are bounded,
it holds that $C_s=O(n_in_ckn)$. Thus, if only $k << n$ columns of the affinity matrix $W$ are computed,
then the total cost $C_t$ is dominated by  $C_k$, i.e., $C_t=O(nm)$.
In the numerical experiments of the next section we show that the immersion based on the farthest triangle provides a good
approximation of $W$ and therefore it is useful to perform the segmentation.

\section{Numerical experiments}\label{NumericalExp}

To prove the performance of the mesh segmentation algorithm proposed in this paper we wrote two main codes. The first code
is the basis for the experiment developed in section \ref{NA1}, which illustrates the advantages and limitations of the
low dimensional embeddings previously considered. The second code is an implementation of the algorithm {\it FSS}
whose results are reported in sections  \ref{NA2} and \ref{NA3}.

We recall that the mesh segmentations shown in this section are computed without including any  procedure to
improve the quality  (smoothness of the boundaries of the segments or their concavity), such as  proposed in \cite{SSCO08},
\cite{WTKL14}.

\subsection{Low dimensional embeddings} \label{NA1}
As we previously mention a valid strategy to solve the segmentation problem is based on computing a low rank approximation
of the affinity matrix $W$. In the following experiment we compare the approximation power of the
low dimensional embeddings described in sections \ref{embbedings} and \ref{algor}. Given a mesh with $n$ triangles  we compute the complete
affinity matrix $W$ of order $n$ given by (\ref{afinidad}). Furthermore, the projection $A^k$ of $W$ on different spaces
for $k=1,...,n$ is also computed. More precisely, given a value of $k$, four projections are computed, each one obtained
when $A^k$ is the matrix $E^k$ given by (\ref{eigmatrix}), the matrix $F^k$ given by (\ref{proyNystrom}), the matrix $G^k$
given by (\ref{proyleverage}) and the matrix $H^k$ given by (\ref{Hk}).

For each  approximation $A^k$, the absolute  error in  Frobenius norm,
\begin{equation}
error_{abs}=\|W-A^k\|_F
\label{errorabsoluto}
\end{equation}
is computed and compared to the error (\ref{mejoraprox}) of the best approximation $E^k$.

\begin{figure}[ht]
\centering
\includegraphics[scale=0.2]{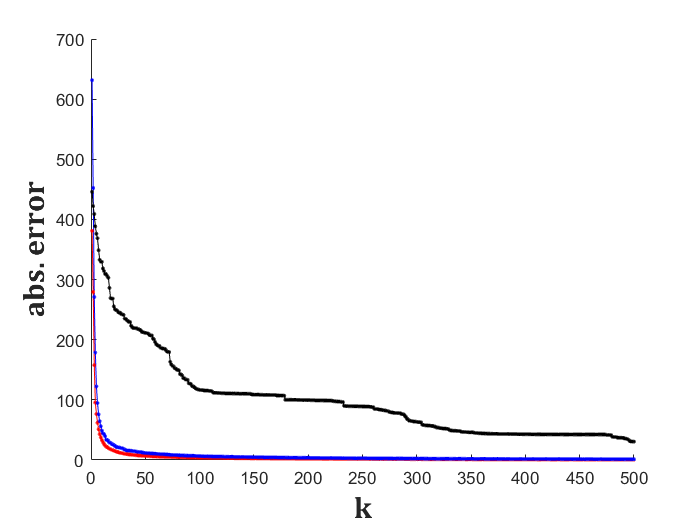}
\includegraphics[scale=0.2]{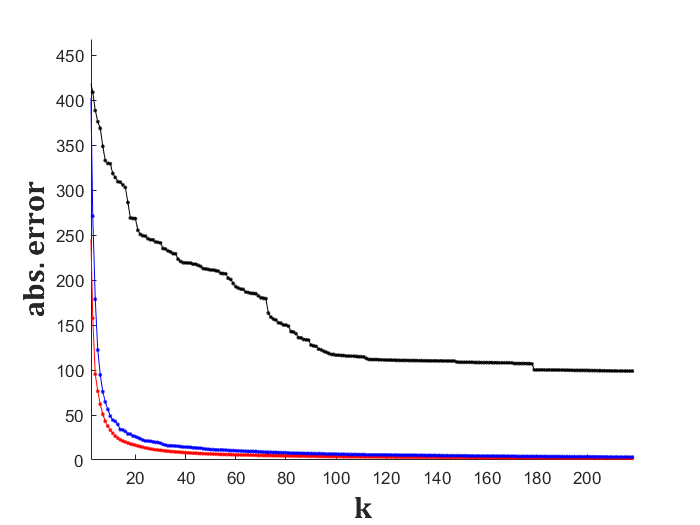}
\includegraphics[scale=0.2]{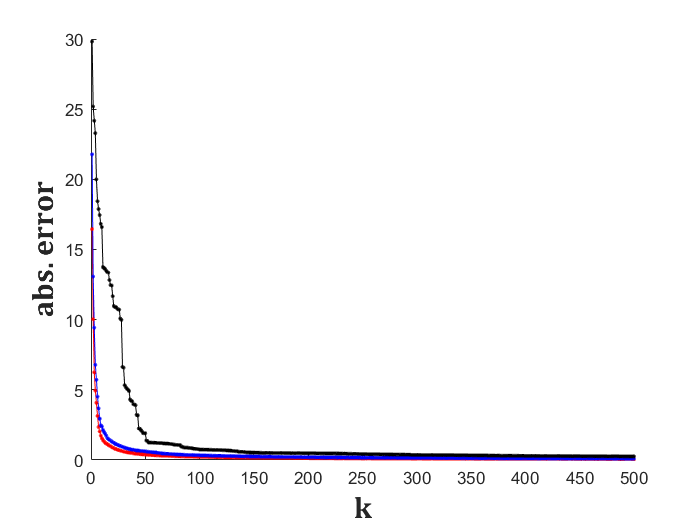}
\includegraphics[scale=0.2]{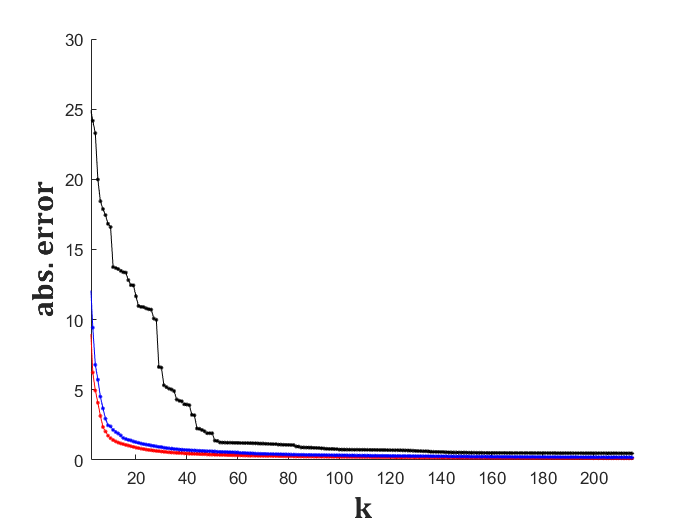}
\caption{Absolute errors (\ref{errorabsoluto}) computed for increasing values of $k$ using different approximations
$A^k$ of the affinity matrix $W$: best (red), leverage (magenta), ours and Nystr\"{o}m (blue). From left to right: errors curves for the
octopus model with 2682 faces, zoom of a section of the first image, error curves for the  hand model with 3000 faces,
zoom of a section of the third image. }
\label{Figabserror}
\end{figure}

\begin{figure}[ht]
\centering
\includegraphics[scale=0.2]{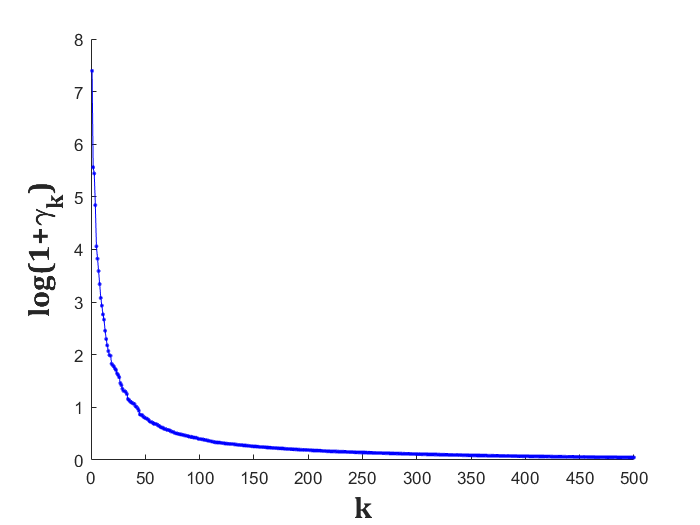}
\includegraphics[scale=0.2]{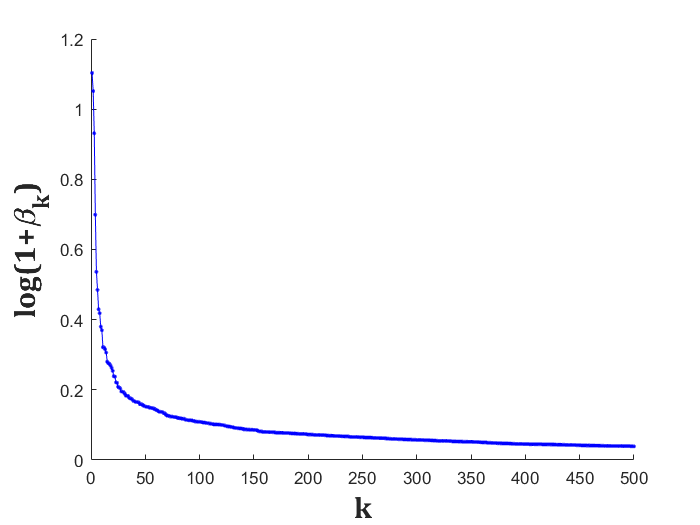}
\includegraphics[scale=0.2]{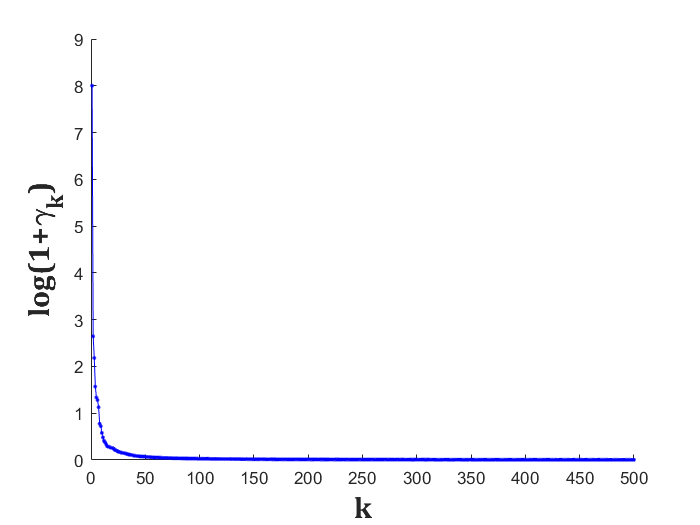}
\includegraphics[scale=0.2]{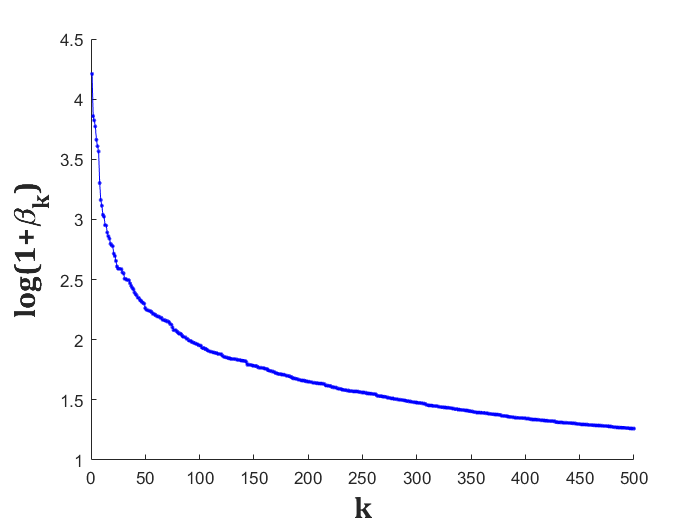}
\caption{The curves $\log(1+\gamma_k)$ and $\log (1+\beta_k)$, both plotted as functions of $k$. From left to right:
the curves for the octopus model with 2682 faces and $k=1,...,500$ and the curves for the hand model with 3000 faces and $k=1,...,500$. }
\label{FigBetaSV}
\end{figure}

Figures \ref{Figabserror} and \ref{FigBetaSV} show the results obtained for two 3D triangulation representing an octopus
and a hand ( models 125.off and 200.off of the Princeton Segmentation Benchmark \cite{CGF09}). In these examples, the distance
matrix $D$ was computed using the {\it geodesic metric} to measure the distance between triangles. In Figure \ref{Figabserror}
we plot the absolute error (\ref{errorabsoluto}) as function of $k$. The red curve corresponds to the error
(\ref{errorabsoluto}) obtained  when $A^k$ is the matrix $E^k$ of the best rank $k$ approximation of $W$. Similarly,
magenta and blue curves are computed with $A^k=G^k,H^k$ respectively.
Since $H^k=F^k$ (see Proposition 1) the curve corresponding to Nystr\"{o}m projection agrees with the blue curve corresponding to
the embedding proposed in this paper. Observe that this curve is the closest to the curve obtained for the embbeding corresponding
to the best approximation of the affinity matrix.

Figure \ref{FigBetaSV} shows curves $\log(1 +\gamma_k)$ and $\log (1+ \beta_k)$, where $\gamma_k$ is the $k$-th
singular value of $W$ (in descending order) and $\beta_k$ is given by (\ref{betak}) (with $l=k$). These curves
correspond to the octopus and the hand models and all decrease very fast.

In general, in all our experiments with several triangulations we observe that:
\begin{enumerate}
\item The absolute error curves show that the matrices $H^k$ and $F^k$ (recall that $H^k=F^k$) provides the approximation to $W$ closest to the optimal $E^k$ for $k \leq n/2$. In practice, we are interested in a good rank $k$ approximation of $W$
with $k << n$. Hence, the embedding corresponding to the farthest point sampling provides the better approximation
with the lower computational cost. It explains experimentally why the mesh segmentation algorithm {\it FSS}
compares favorably to another methods reported in the literature, which are based on the spectrum of $W$ and happen
to be more expensive.
\item The method proposed in this paper  cannot be considered as a random method, since except the first column of the sample,
the rest of the columns are selected deterministically. However, one may think of  $1 - \beta_l/\beta_1,\;l > 1$ as the
{\it conditional probability} of selecting the $j_l$-th column of $W$ given that the columns $j_1,...,j_{l-1}$ have been
previously selected. In this context, our farthest point sampling scheme may be considered as an algorithm to select the $k$
columns with highest conditional probability (see more details in section \ref{section5.4}).
\end{enumerate}

\subsection{ Mesh segmentation qualitative performance}\label{NA2}
In this section we show the performance of our mesh segmentation algorithm {\it FSS} with several 3D triangulation models.
In the experiments reported in this section and in the next one, the {\it kmeans ++} \cite{AV07} algorithm is applied to the
normalized  rows of the rectangular matrix $W^k$, composed by the $k$ selected columns of $W$. These rows
are considered as points in $\mathbb{S}^{k-1}$. To measure the distance between two vectors we use the {\it cosine distance}, which
means that the distance is defined as one minus the cosine of the included angle between them. Several replicates of {\it kmeans ++}
algorithm are applied and for each replicate  the {\it seeds} of the $n_c$ clusters are selected randomly. In general, the algorithm
{\it FSS} works very fast since in all segmentations $k$  is at most $10\%$ of the total number of faces. Our goal here
is just to evaluate {\it visually} the quality of the segmentations produced by the algorithm.

\begin{figure}[ht]
\centering
\includegraphics[scale=0.16]{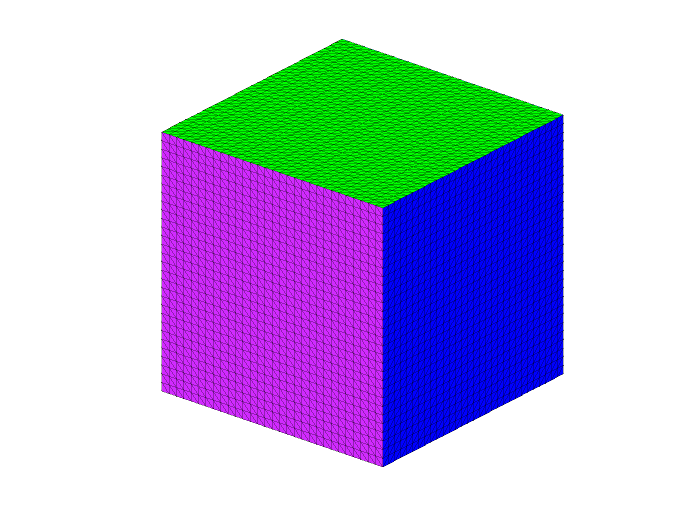}\hspace{1cm}
\includegraphics[scale=0.2]{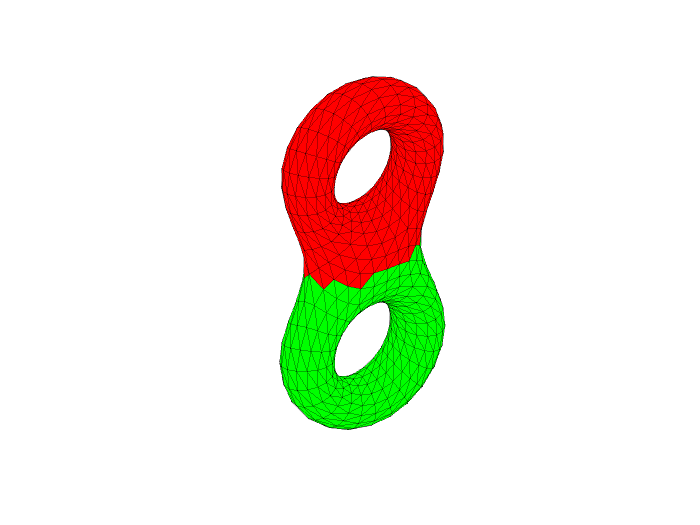}\hspace{0.6cm}
\includegraphics[scale=0.23]{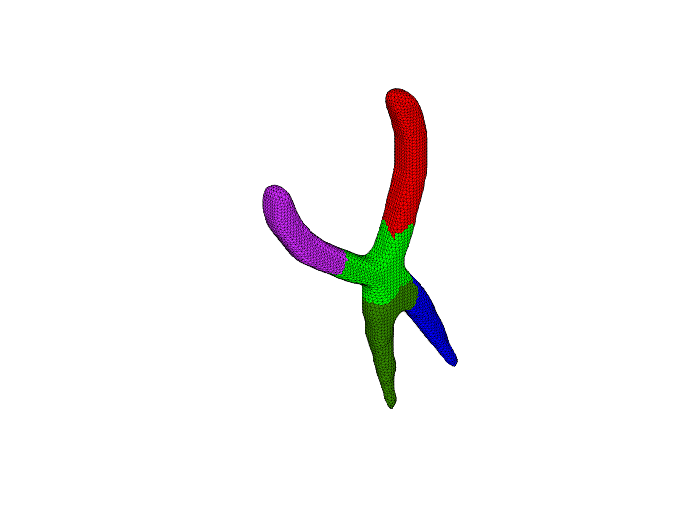}
\caption{Left: segmentation with the {\it angular distance} of the cube model. The clusters are obtained computing $1\%$ of
 the $10800$ columns of $W$. Center: segmentation of the eight model  with the {\it geodesic distance} obtained computing
 $2\%$ of the $1536$ columns of $W$. Right: pliers model segmentation with the {\it sdf distance} obtained computing $5\%$ of
 the $8970$ columns of $W$.}
\label{Fig:cube}
\end{figure}

The first example that we have considered is the model of a cube defined by a triangulation with $10\,800$ faces. This model is
ideal to check how the algorithm works when the distance between triangles is measured in terms of the {\it angular distance}.
To segment the model we computed only  $1\%$ of the columns of the affinity matrix $W$.  Figure \ref{Fig:cube} left shows that the
results are excellent, since the faces of the cube correspond exactly with the 6 clusters produced by the automatic segmentation.
In the second example, we use the {\it geodesic distance} between triangles to define the affinity matrix $W$ of the eight model.
In Figure \ref{Fig:cube} center we show the segmentation in $2$ clusters of the model, obtained computing only $2\%$ of the
columns of $W$. Observe that each cluster agrees approximately with one \textit{handle} of the eight. In our
third example, we use the {\it sdf distance}  to segment the pliers model in $5$ clusters. In  Figure \ref{Fig:cube} right we show
the results obtained computing $5\%$ of the columns of $W$. As in the previous examples, the clusters are natural partitions of the model.

\begin{figure}[ht]
\centering
\includegraphics[scale=0.3]{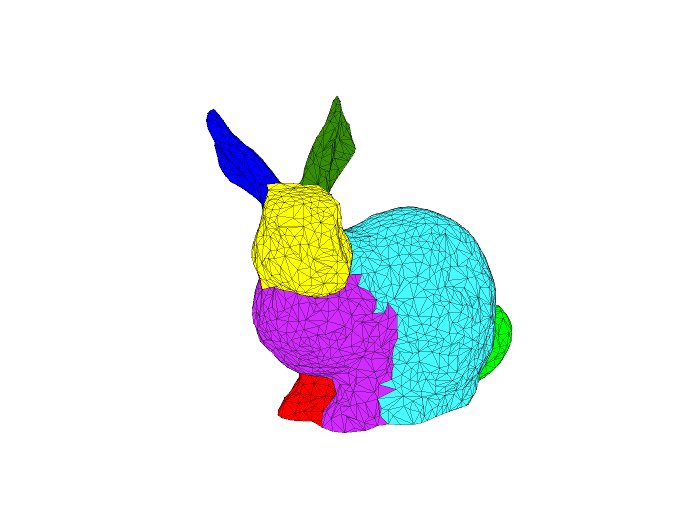}
\includegraphics[scale=0.3]{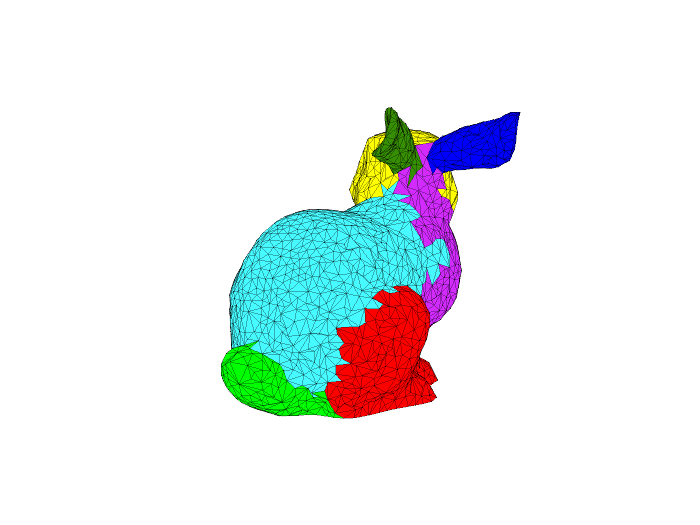}\\
\includegraphics[scale=0.3]{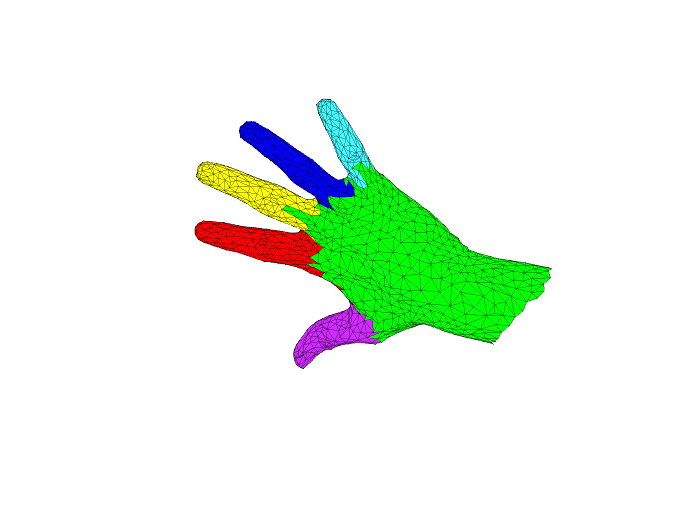}
\includegraphics[scale=0.3]{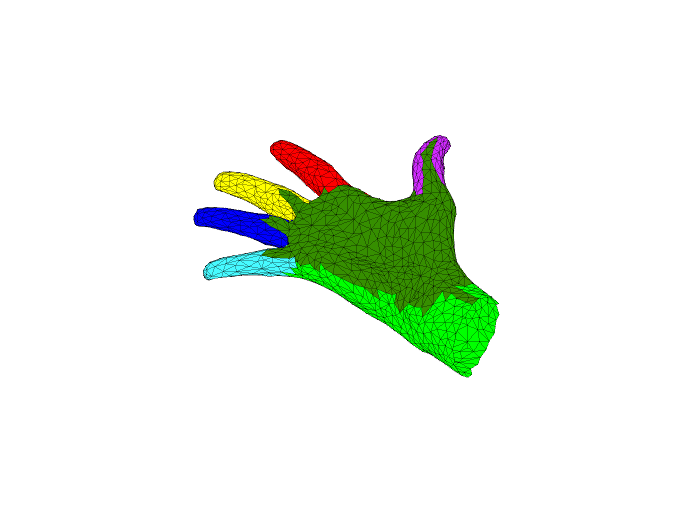}
\caption{Two segmentations obtained with the metric defined as the product of geodesic and angular distances. Top: two
views of the segmentation of the bunny model. The segmentation was computed using $10\%$ of the $3860$ columns of the
affinity matrix $W$. Bottom: two view of the segmentation of the hand model, segmentation obtained computing  $1\%$ of
the $3000$ columns of $W$  .}
\label{Fig:bunny2k2}
\end{figure}

In our next examples we use a product metric to compute the distance between neighboring triangles. More precisely,
if the faces $f_i$ and $f_j$ share an edge of the triangulation, then the {\it product distance} $d_{ij}$ between them is
defined as $d_{ij}:=d_{ij}^gd_{ij}^a$, where $d_{ij}^g$, $d_{ij}^a$ are the geodesic and angular distances between $f_i$
and $f_j$ respectively. As usual, the distance between no adjacent faces is defined as the length of the shortest path
in the dual graph. Figure \ref{Fig:bunny2k2} left shows the segmentation in $8$ clusters of the bunny model. This result
 was obtained computing the $10\%$ of the total number of  columns of the affinity matrix $W$. Observe that the segmentation
 produced by the product distance distinguishes well not only the big ears but also the small tail.
The hand model  with $3000$ faces is more challenging. As we observe in Figure  \ref{Fig:bunny2k2} right, there is some
leakage in the clusters corresponding to the fingers, even when this leakage is substantially smaller than the one shown
in the hand model in \cite{LZ04}. Moreover, the palm and the back of the hand belong to different clusters since the combined
metric is not enough to capture all the volumetric information. These limitations could be overcome if we include in the
definition of the combined metric a part-aware distance, such as the part-aware distance in \cite{LZSC09}.

\subsection{ Mesh segmentation quantitative performance}\label{NA3}

In this section we study the behavior of our mesh segmentation algorithm {\it FSS} through several examples
of the Princeton Segmentation Benchmark \cite{CGF09} for evaluation of 3D mesh segmentation algorithms. This benchmark
comprises a data set with 380 surface meshes of 19 different object categories. It also provides a ground-truth corpus
of 4300 human segmentation.

In the next examples we observe that the algorithm proposed in this paper, based on the computation of {\it few} columns
of the affinity matrix, produces segmentations  which {\it compare} to the results obtained using the {\it full} affinity
matrix.  Recall that it doesn't mean that the segmentation obtained with few columns of the affinity matrix $W$ is always
good, but that it is as good as the one obtained computing all columns of $W$. In other words, if we select carefully which
columns of $W$ are computed, then the quality of the results essentially depends on how good the selected metric reflects
the features of the triangulation. In our experiments we compute the distance between triangles using several metrics:
the \textit{geodesic} distance, the \textit{angular} distance, the product of them and the \textit{sdf} distance \cite{SSCO08}.

Usually, the quantitative evaluation of a segmentation algorithm is done by comparing the automatic segmentation with one
or more reference segmentations of the ground-truth corpus. In the literature one can find several metrics to evaluate
{\it quantitatively} the similarity between two segmentations of a triangulated surface \cite{BVLD0}, \cite{CGF09}.
In this section  we employ two different non-parametric measures: the Jaccard index $JI$ \cite{FM83} and the Rand index
$RI$ \cite{Rand71}. The segmentation of a mesh with $n$ triangles may be described by a vector $s=(s_1,...,s_n)$,
where $s_j$ is the index of the cluster to which the $j$-th triangle belongs. Given two segmentations $s_a$ and $s_b$
of the same mesh, we denote by $JI(s_a,s_b)$ and $RI(s_a,s_b)$ the similarity between them according to the Jaccard
and Rand indexes respectively. For both indexes it holds: $0 \leq JI(s_a,s_b) \leq 1$ and $0 \leq RI(s_a,s_b) \leq 1$,
where the value $1$ corresponds to the maximal similarity, i.e. $JI(s_a,s_b)=1$ or $RI(s_a,s_b)=1$ means that segmentations
$s_a$ and $s_b$ are identical. In the experiments we compute Jaccard and Rand distances between $s_a$ and $s_b$ given
by, $d_J(s_a,s_b):=1-JI(s_a,s_b)$ and $d_R(s_a,s_b):=1-RI(s_a,s_b)$.

\begin{figure}[htb]
\centering
\includegraphics[scale=0.22]{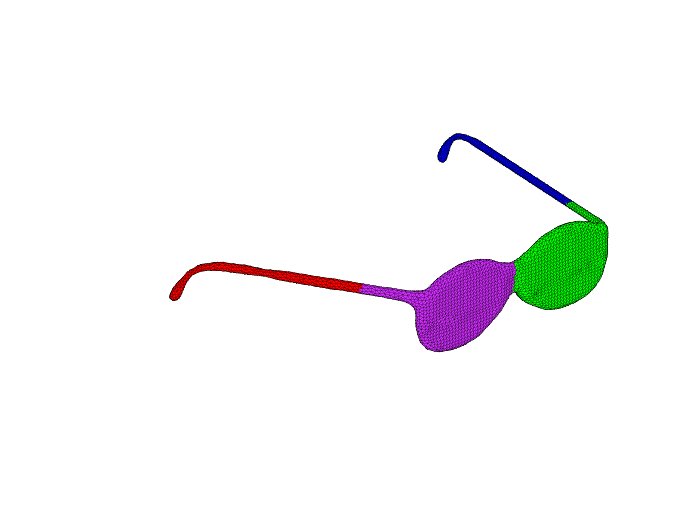}
\includegraphics[scale=0.22]{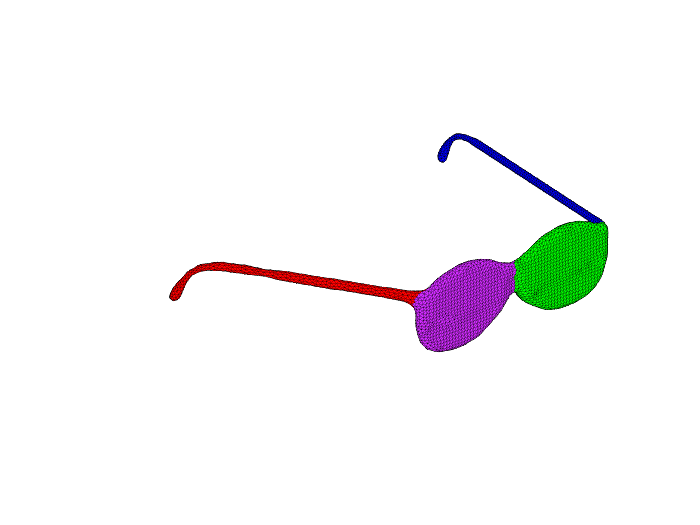}\\
\includegraphics[scale=0.22]{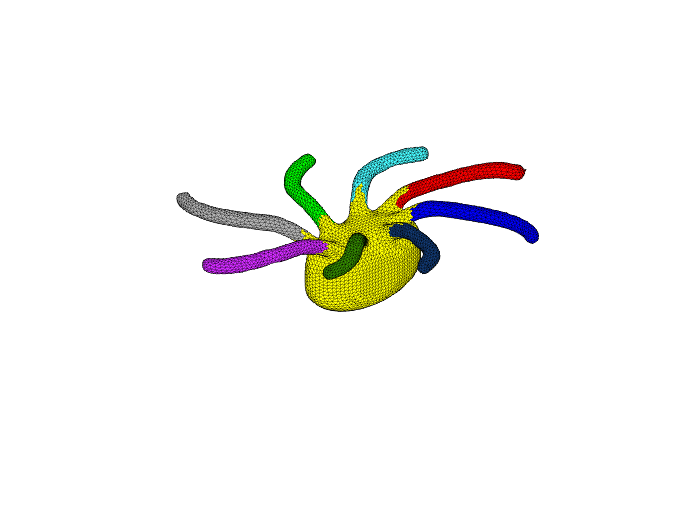}
\includegraphics[scale=0.22]{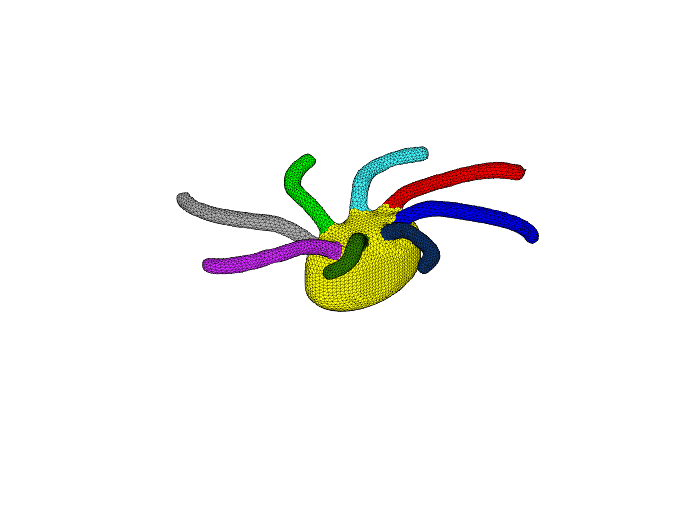}\\
\includegraphics[scale=0.22]{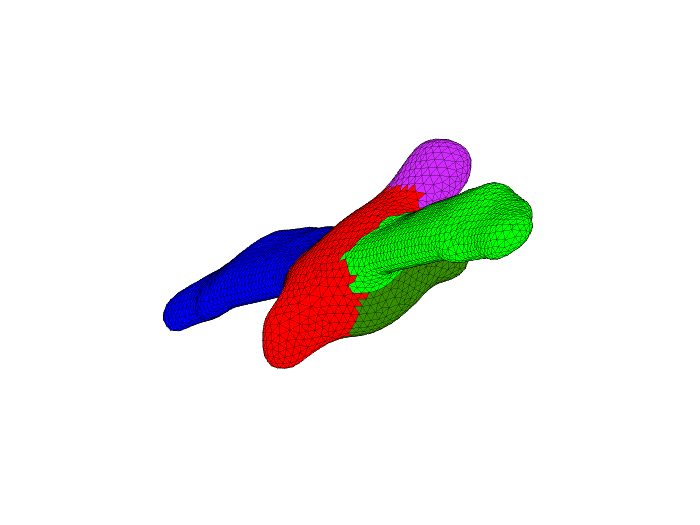}
\includegraphics[scale=0.22]{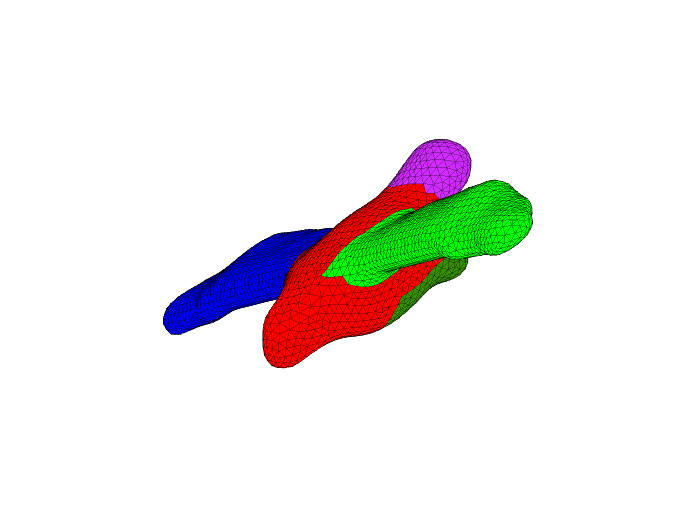}
\caption{ Left column: segmentations obtained computing few columns of the matrix $W$, right column: ground truth segmentations.
First row: segmentation  of the sunglasses based on \textit{geodesic} distance and obtained computing $2\%$ of the $8324$
columns of $W$. Second row: segmentation  of the octopus based on the \textit{angular} distance and obtained computing $1\%$
of the $11888$ columns of $W$. Third row: segmentation of the bird based on the \textit{sdf} distance and obtained computing
$1\%$ of the $6312$ columns of $W$.}
\label{Fig:gafaspulpomujer}
\end{figure}

In Figure \ref{Fig:gafaspulpomujer} we show the results obtained for three models of the  Princeton Segmentation
Benchmark: the sunglasses (model 42), the octopus (model 121) and the bird ( model 243). For all models, Rand and
Jaccard distances are computed comparing the automatic segmentation (left column) with and the ground truth segmentation
(right column). Table 1 below shows the values of Rand and Jaccard distances as well as the percent of columns
of the affinity matrix used to obtain the automatic segmentation and the metric employed for computing the distance
between triangles. In these examples a low percent of columns provides good segmentations.\\

\begin{table}[ht]
\centering
\begin{tabular}{|c|c|c|c|c|}
\hline
Example & distance & $\%$    & $d_R$    & $d_J$  \\
\hline
  glasses & geodesic & $2\%$   & $0.062$  & $0.161$ \\
  octopus & angular  & $1\%$   & $0.052$  & $0.100$ \\
  bird   & sdf  & $1\%$   & $0.048$  & $0.295$ \\
\hline
\end{tabular}
\caption{ Rand and Jaccard distances ( $d_R$ and $d_J$, respectively) between the automatic segmentation and
the ground truth segmentation.}
\end{table}

\begin{figure}[bht]
\centering
\includegraphics[scale=0.18]{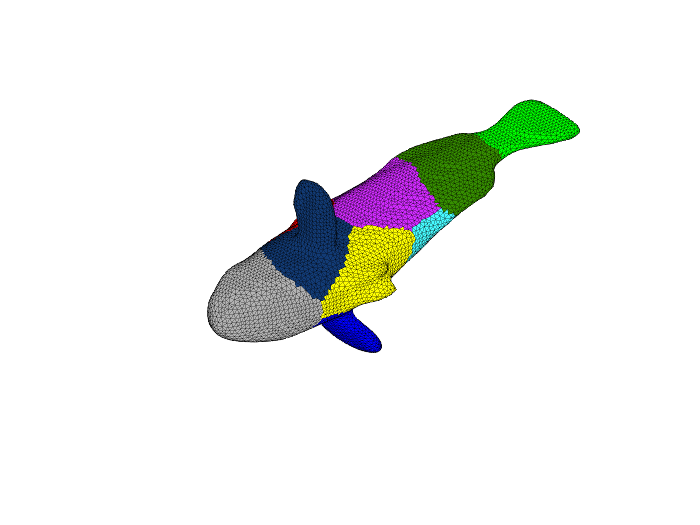}
\includegraphics[scale=0.18]{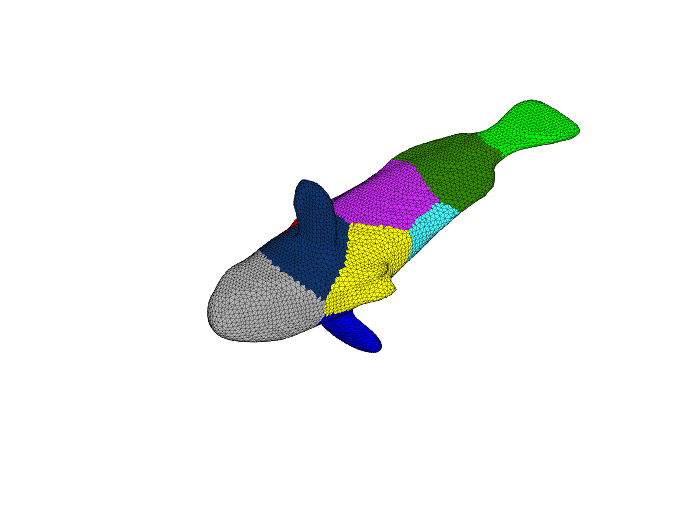}
\includegraphics[scale=0.18]{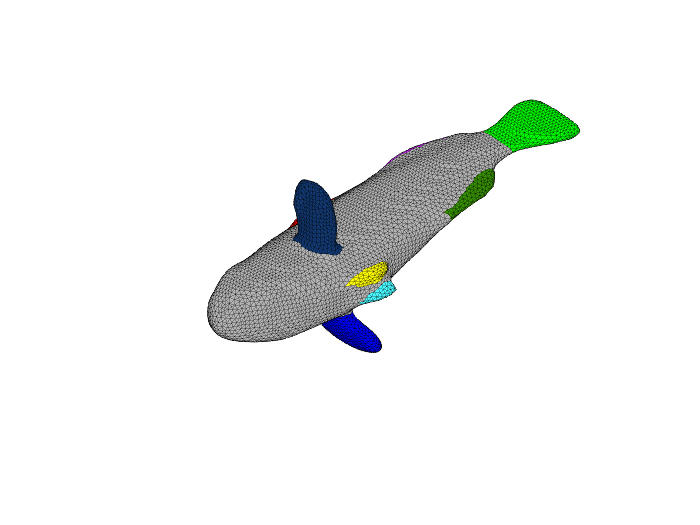}
\caption{Three segmentations of a fish (model 225 of the Princeton Segmentation Benchmark). Left and center: segmentations
based on the \textit{geodesic} distance obtained using the $0.5\%$ and  the $100\%$ of the $12148$ faces. Rand and Jaccard distances
between them are $d_R=0.010$ and $d_J=0.064$ respectively. Right: ground truth segmentation. Rand and Jaccard distances
between the segmentation obtained computing all the columns of $W$ and the ground truth segmentation are $d_R=0.415$, $d_J=0.742$.}
\label{Fig:fish}
\end{figure}

In Figure \ref{Fig:fish} we illustrate that the quality of the automatic segmentation with few columns of $W$ strongly depends
on the capability of the selected metric to capture the features of the mesh. In this example we show that even if we compute
 the whole affinity matrix $W$, the resulting automatic segmentation is far a way from the ground truth segmentation,
 as it is shown by the values of $d_R$ and $d_J$ between these segmentations ($d_R = 0.415$, $d_J = 0.742$). On the other
 hand, the automatic segmentations obtained computing $0.5\%$ and $100\%$ of the columns of $W$ are very similar, since the
 values of $d_R$ and $d_J$ between them are small ($d_R=0.010$ and $d_J=0.064$). Hence, the segmentation with few columns
 is also not good in comparison with the ground truth segmentation. Finally, in this example we also observe that the body
 of fish is subdivided in clusters with similar areas. As other authors pointed out \cite{CGF09}, this behavior is typical
 of segmentations based on {\it k-means} algorithm.

\subsection{Quality of segmentations: a different way of measuring.}\label{section5.4}
In the experiments of this section we use a different approach to measure the quality of the segmentation. Instead of comparing
the automatic segmentation, obtained computing $k$ columns of the affinity matrix,  with the ground-truth of the corpus,
we compare it with the segmentation obtained using {\it all} columns of the affinity matrix. In our opinion, this
comparison is fairer, since the simple metrics (\textit{geodesic}, \textit{angular} and \textit{sdf} distances) that we have used to compute the distance and affinity matrices are not always enough to produce good segmentations. Hence, the comparison of the automatic
segmentations with ground-truth corpus segmentations does not help in the sense of proving that the results with few
{\it well} selected columns of the affinity matrix are of quality quite similar to the results obtained computing all
affinity matrix.\\

\begin{figure}[htb]
\centering
\includegraphics[scale=0.25]{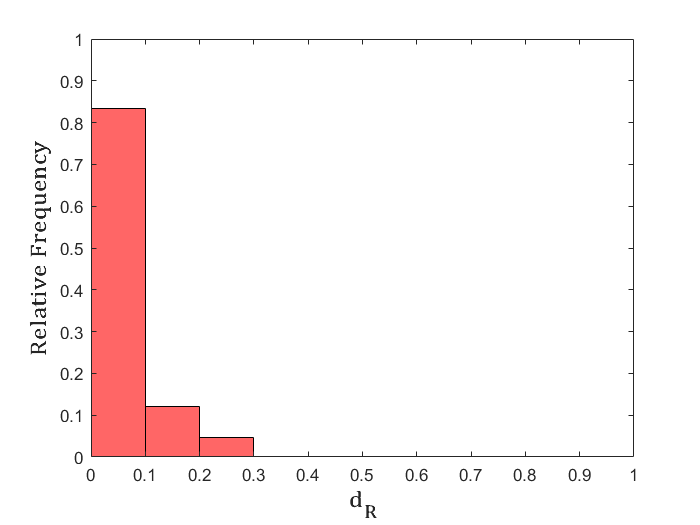}
\includegraphics[scale=0.25]{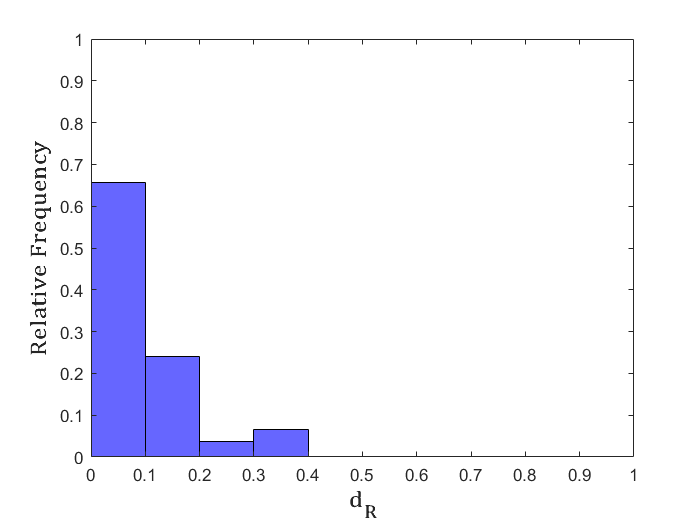}
\includegraphics[scale=0.25]{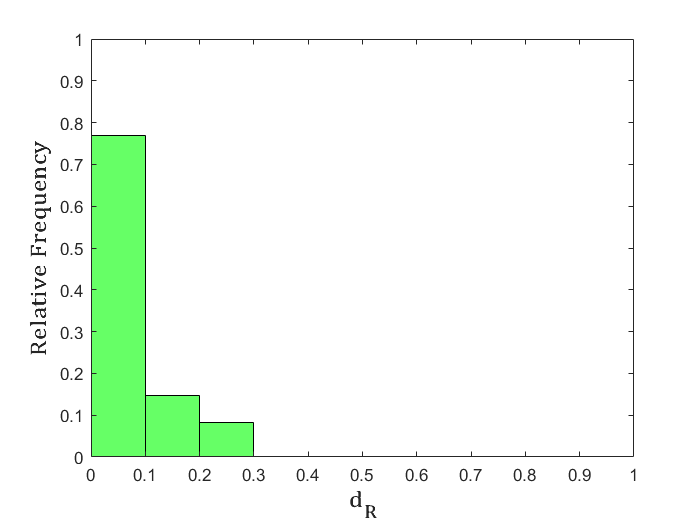}
\caption{Relative frequency histograms of the Rand distance values, $d_R$,  between the segmentations obtained with the whole  matrix $W$
and the segmentation  obtained with the same metric for $k$ columns of $W$, with $k=0.5\%,1\%,2\%,5\%,10\%,25\%$. From left to right: \textit{geodesic},\textit{angular} and \textit{sdf} metric.}
\label{Fig:barras}
\end{figure}

Applying our segmentation method,  $18$  meshes of the Princeton Segmentation Benchmark \cite{CGF09} are segmented
using $k$ columns of the affinity matrix $W$, where $k=0.5\%,1\%,2\%,5\%,\%10$ and $\%25$ of the total number $n$ of faces.
The Rand distance $d_R$ between the segmentations obtained for $k$ columns  and the segmentation  obtained with the
same distance for the whole  matrix $W$ is computed. Three different metrics are considered, the \textit{geodesic},
the \textit{angular} and the \textit{sdf} metrics. For each of these metrics, in Figure \ref{Fig:barras} it is shown
the  histogram of relative frequency of the Rand distance values between the segmentations obtained for $k$ columns
and the segmentation  obtained with the same metric for the whole  matrix $W$. This experiment shows that with
high probability, the segmentations with small number of columns $k$ are very close to the corresponding
segmentation  obtained  for the whole  matrix $W$.\\

\subsection{The $\beta$ curve}\label{beta}

\begin{figure}[htb]
\centering
\includegraphics[scale=0.25]{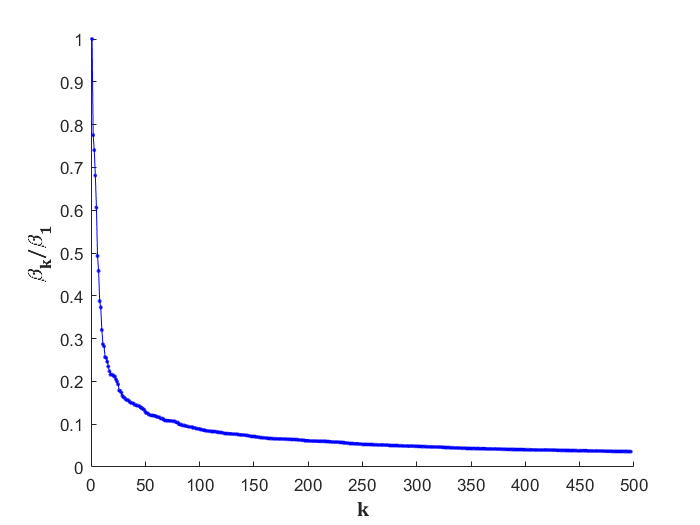}
\includegraphics[scale=0.25]{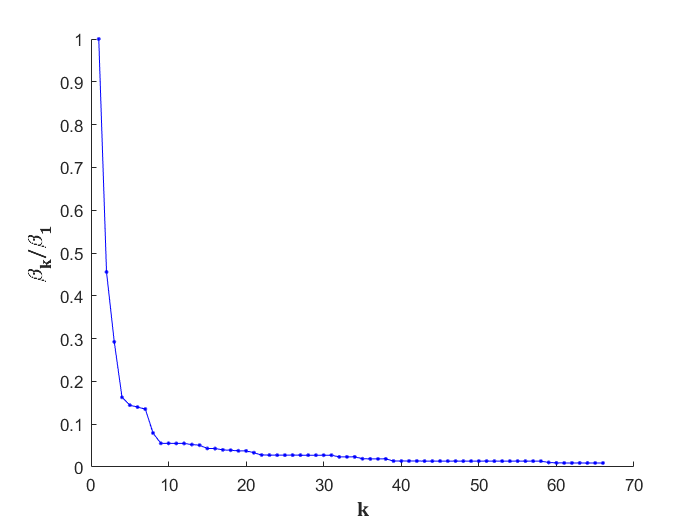}
\includegraphics[scale=0.25]{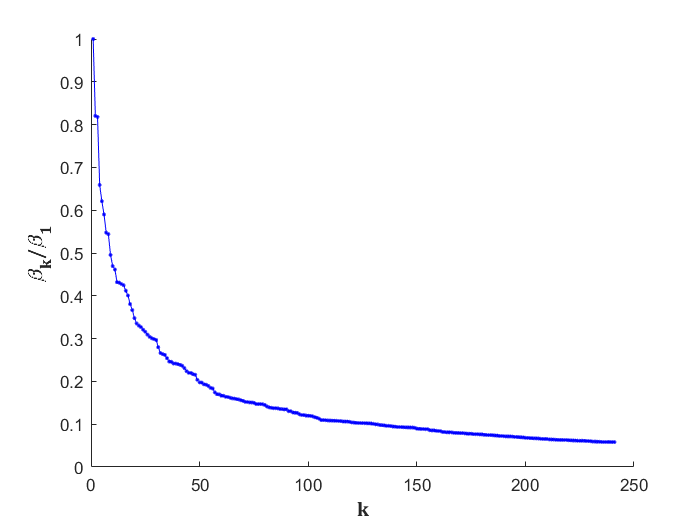}
\includegraphics[scale=0.25]{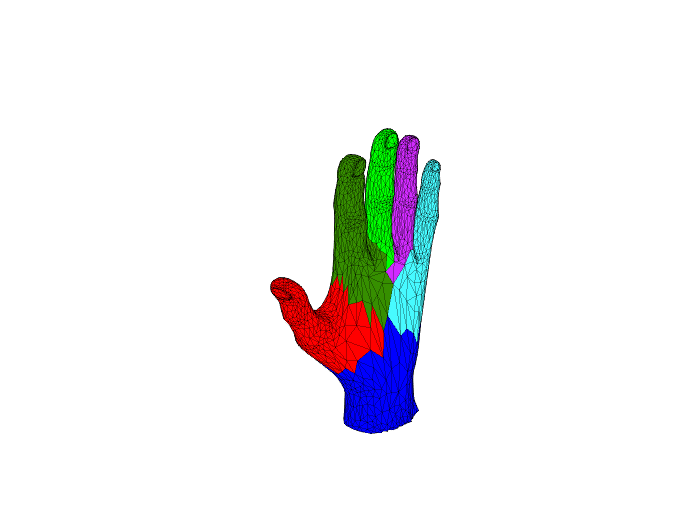}
\includegraphics[scale=0.25]{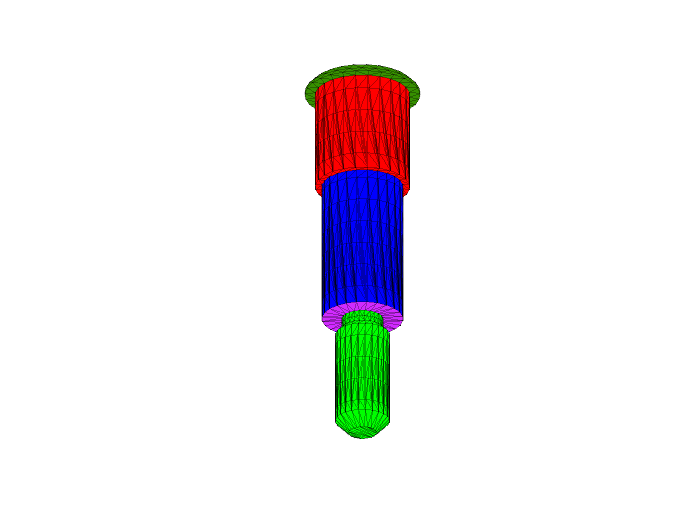}
\includegraphics[scale=0.25]{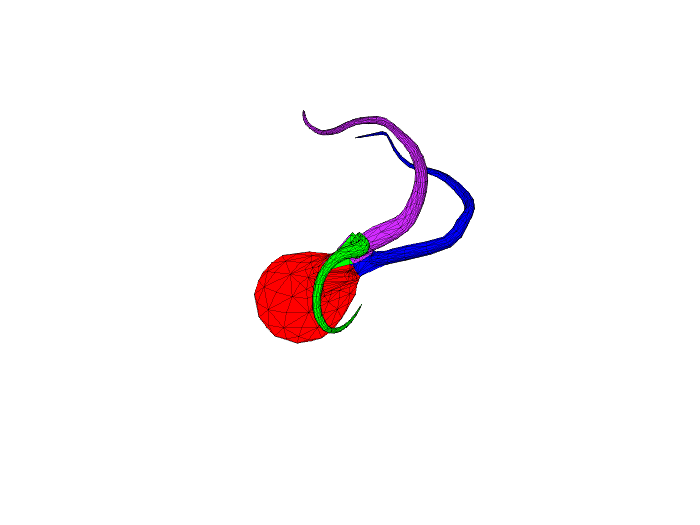}

\caption{First row: normalized curve $\beta_k/\beta_1$ curve. Second row: segmentation of the models with $k^{*}$ columns of $W$.}
\label{Fig:curvabeta185}
\end{figure}

The graph of $\beta_k$ (\ref{betak}) as function of $k$ has a ``{\it L}"-shape, similar to the singular values curve,
which decreases very fast for small values of $k$, see Figure \ref{FigBetaSV}. It suggests that $\beta$ curve could
be used to propose an lower bound for the size $k$ of the sample that furnishes a projection $H^k$ providing a
good approximation to $W$. In the first row of Figure \ref{Fig:curvabeta185} it is shown a section of the $\beta$ curve
$(k,\frac{\beta_k}{\beta_1}),\;1 \leq k \leq n$ for three models of the Princeton Segmentation Benchmark (hand, bearing, octopus),
considering different distances (\textit{geodesic, angular, sdf}, respectively). Observe that these curves decrease
very fast for small values of $k$ and tend slowly to 0 when $k$ goes to $n$.

For several values of $k$, Table \ref{tablitza} shows  the Rand and Jaccard distances between the automatic segmentation
 $s_k$ and the ground truth segmentation of the Princeton Segmentation Benchmark. The smallest number of columns
 for which the slope of the $\beta$ curve may be considered as very small are marked with $*$. Observe that for any fixed model
 and all considered values of $k$, the values of the Rand and Jaccard distances between the automatic segmentation and
 the ground truth segmentation are very similar. In the second row of Figure \ref{Fig:curvabeta185} the segmentations
corresponding to $k^*$ are shown.

\begin{table}[ht]
\centering
\begin{tabular}{|c|c|c|c|c|}
  \hline
  Example & distance           & $k$       & $d_R$     & $d_J$  \\
  \hline
  hand    & \textit{geodesic}  & $49^* $   & $ 0.124$  & $ 0.303$ \\
   $4979$ faces      &         & $497 $    & $0.124 $  & $0.303 $ \\
                     &         & $1243 $   & $0.123 $  & $ 0.302$ \\
   \hline
  bearing & \textit{angular}   & $16^*$    & $0.033 $  & $ 0.452$ \\
     $3322$ faces     &        & $66 $     & $0.033 $  & $ 0.452$ \\
                      &        & $166 $    & $0.028 $  & $ 0.452$ \\
    \hline
  octopus & \textit{sdf}       & $53^*$    & $0.043$   & $ 0.105$ \\
  $2682$ faces        &        & $187 $    & $0.040$   & $ 0.100$ \\
                      &        & $321 $    & $0.040$   & $ 0.099$ \\
\hline
\end{tabular}
\caption{ Rand and Jaccard distances ( $d_R$ and $d_J$ respectively) between the proposed automatic segmentation and the
ground truth segmentation. The automatic segmentation was obtained using the metric indicated in the second column and
computing the number of columns of the affinity matrix indicated in the third column.}
\label{tablitza}
\end{table}

 \subsection{Comparing our method with the spectral approach}
 Since the method {\it FSS} proposed in this paper has some points of contact with the one introduced in \cite{LZ04} and improved in \cite{LJZ06}, in this section we compare the segmentations obtained with both approaches. The algorithm introduced in \cite{LJZ06} applies
 Nystr\"om's method to approximate the spectral embeddings of faces of the triangulation. To avoid the expensive computation
 of the normalized matrix $Q=M^{-1/2}WM^{-1/2}$ and its largest eigenvectors,  Nystr\"om's method  computes approximately
 the largest eigenvectors of $Q$, from a  small sample of its rows (or columns) and the solution of a small scale
 eigenvalue problem (see section \ref{sectionAm}). The final step consists in applying \textit{k-means} to the rows of $Q$. The selection of the sample has a strong influence  on the accuracy of the approximated eigenvectors.

 No comments on the recommended relationship among the size of the sample and the number of eigenvectors are included in \cite{LJZ06}. In the numerical experiments reported here, the same sample of max-min farthest faces  is used to select the columns of $W$ to be computed by our method {\it FSS} and also for the Nystr\"om approximation of the largest eigenvectors of $W$. Further, in the results obtained with the spectral method, we set the number of  eigenvectors equal to the number of clusters. Moreover, as suggested in \cite{FBCM04}, the Nystr\"om approximated eigenvectors of $Q$ are orthogonalized before applying {\it k-means} clustering.

\begin{figure} \centering
 \includegraphics[scale=0.2]{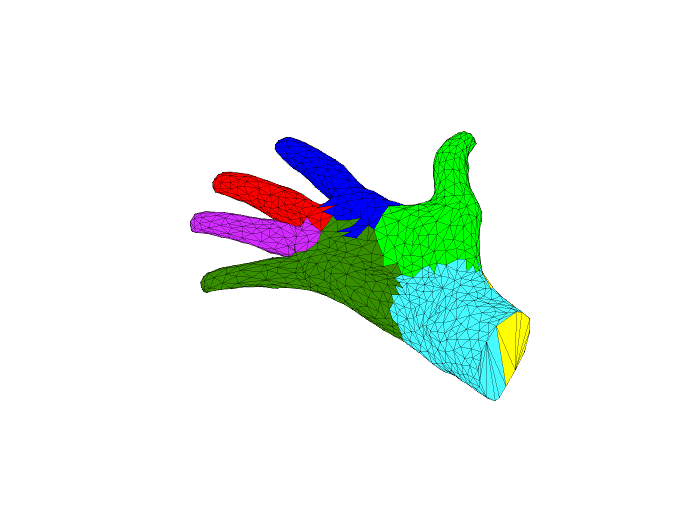}
 \includegraphics[scale=0.2]{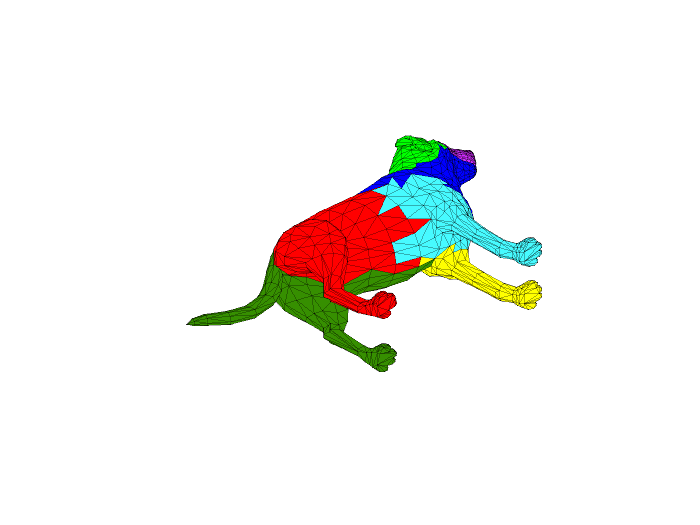}
 \includegraphics[scale=0.23]{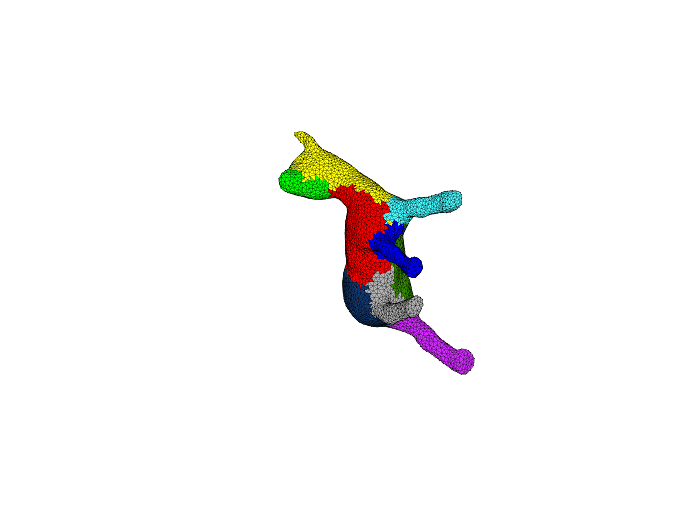}\\
 \includegraphics[scale=0.2]{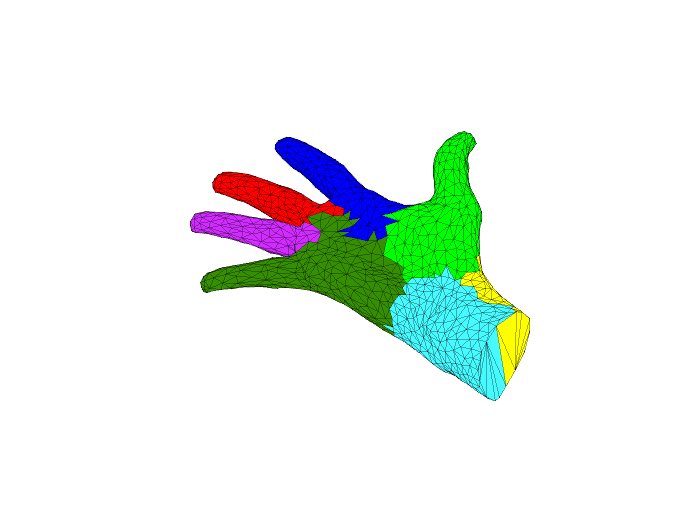}
 \includegraphics[scale=0.2]{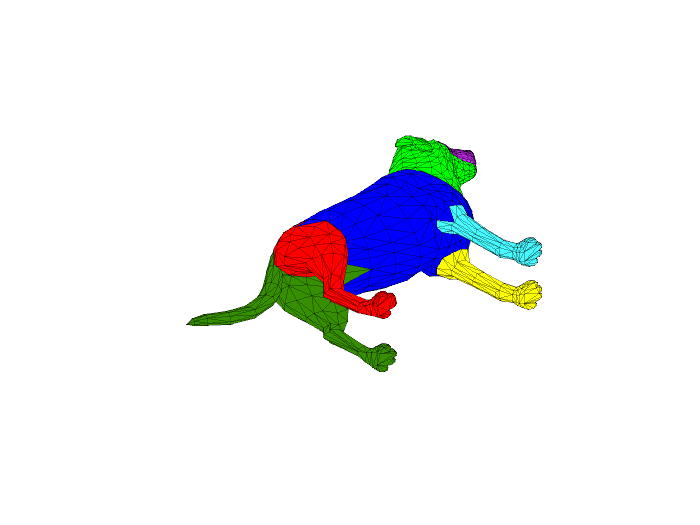}
 \includegraphics[scale=0.23]{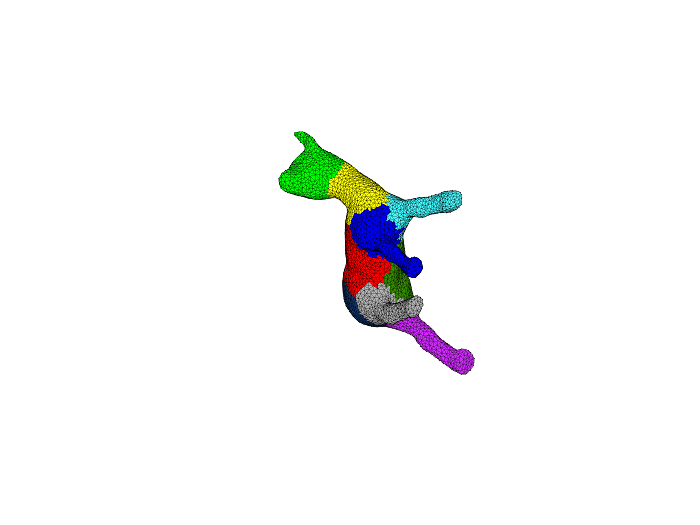}
 \caption{First row: Spectral segmentations with  Nystr\"om approximation, the number of eigenvectors is equal to the number of clusters.
 Second row: segmentations obtained with our method {\it FSS}. The sample of columns of the affinity matrix is the same in both
 approaches: $0.5\%$ for the hand, $25\%$ for dog and $5\%$ for fawn.}
 \label{Fig:Nys}
 \end{figure}

 Figure \ref{Fig:Nys} shows the segmentation based on the {\it sdf distance} of several models,  using the
 spectral method with Nystr\"om approximation and our method {\it FSS}. Both methods compute the same percent of columns of $W$
 corresponding to the farthest triangles. Table 3 shows the values of Rand and Jaccard distances between the corresponding segmentations. In general we observe that even when the segmentation are different, the quality of them is similar. The Rand distance between segmentations are very small in all cases, but Jaccard distances are larger reflecting better the visual differences. \\

\begin{figure}[h!]
\centering
\includegraphics[scale=0.24]{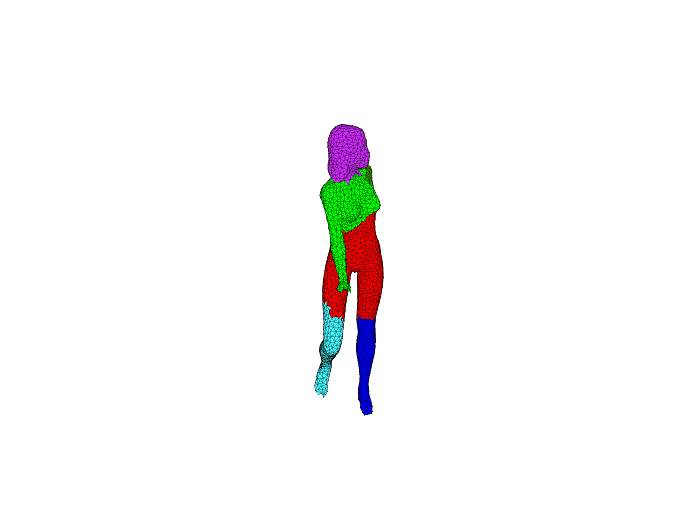}
\includegraphics[scale=0.24]{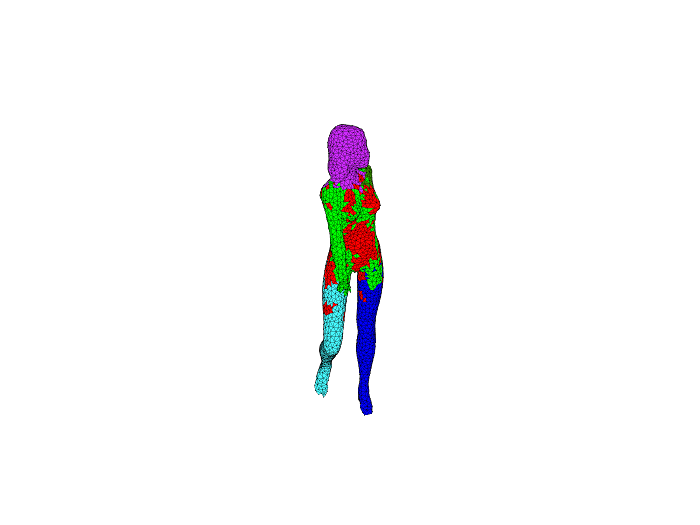}
\includegraphics[scale=0.24]{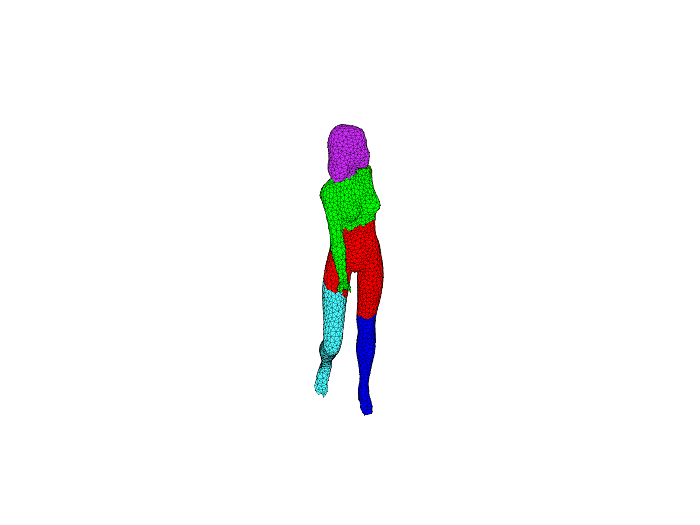}
\caption{ Left: spectral segmentation computing {\it all} columns of the affinity matrix, center: spectral segmentation with
 Nystr\"om approximation computing $5\%$ of columns of the affinity matrix $W$, right: our segmentation based in the
 same sample of columns of $W$.   }
\label{Fig:Compara}
\end{figure}

 \begin{table}[ht]
 \centering
 \begin{tabular}{|c|c|c|c|c|}
   \hline
   Example & $\%$    & $d_R$    & $d_J$  \\
   \hline
   hand   & $0.5\%$  & $0.054$  & $0.229$ \\
    dog   & $25\%$   & $0.082$  & $0.406$ \\
   fawn   & $5\%$    & $0.084$  & $0.397$ \\
 \hline
 \end{tabular}
 \caption{Rand and Jaccard distances between our segmentation and the segmentation produced by the spectral approach
 using  Nystr\"om approximation. The sample of columns of $W$ is the same for both methods. The second column of the table
 indicates the size of the sample which is the indicated $\%$ of the total number of triangles.}
 \end{table}
In our last example we show an unexpected artefact that we have observed. Sometimes the spectral segmentation obtained
using  Nystr\"om approximation produces clusters which are {\it non connected}. This undesirable effect is eliminated
when we increase the size of the sample of columns of the affinity matrix $W$. In contrast, our segmentation approach
using the same sample of columns of  $W$ always produces connected clusters  (recall lemma \ref{polarisation}). In Figure
\ref{Fig:Compara} left we show the spectral segmentation of the woman model obtained with the normalized matrix $Q$.
Center and right images of this figure show the spectral segmentation with Nystr\"om approximation and with the method
proposed in this paper, both using $5\%$ of columns of $W$ corresponding to the farthest triangles in the {\it sdf}
distance. The described artefact becomes evident in the spectral segmentation with Nystrom approximation. \\

\section{Conclusions and future work}

We have proposed a segmentation method for triangulated surfaces that only depends on
a metric to quantify the distance between triangles and on the selection of a sample of
few triangles. The proposed method computes the weighted dual graph of the triangulation
with  weights equal to the distance between neighboring triangles. The $k$ farthest triangles in the
chosen metric are used to compute a rectangular affinity matrix $W^k$ of order $n \times k$, where
the number $k$ of columns is much smaller than the total number of triangles $n$.  Rows of $W^k$
encode the similarity between all triangles and the $k$ triangles of the sample.
Thus, clustering the rows of $W^k$ happens to be consistent with the results of clustering the rows of the whole matrix $W$ and
no artefact appears. Hence, a valid strategy to solve the segmentation problem consists in
clustering the rows of $W^k$ by using for instance the {\it  k-means} algorithm.

From the theoretical point of view the problem of reducing the dimensionality for clustering is strongly
connected with the low rank approximation of the matrix containing the data to be clustered, which in our
context is the affinity matrix $W$. In this sense, we have proved that for any sample $\mathcal{X}$ of $k$ indexes, the
rank $k$ approximation of $W$ obtained projecting it on the space generated by the columns of $W$ with
indexes in $\mathcal{X}$, coincides with the rank $k$ approximation obtained projecting $W$ on
the space generated by its approximated eigenvectors, computed by Nystr\"om's method with the same
sample of columns of $W$.
Moreover, it is shown that if the columns of $W^k$ correspond to the $k$ farthest triangles in the selected metric, then the proximity relationship among the rows of $W^k$ tends to faithfully reflect the proximity among the corresponding rows of $W$.

In practice, our experiments have confirmed that this occurs even for relatively small $k$, resulting in a low computational
cost method. Multiple experiments with a large variety of 3D triangular meshes were performed and they have shown that
the segmentations obtained computing less than the $10\%$ of columns of the affinity matrix $W$ are as good as those
obtained from clustering the rows of the full matrix $W$. We have also observed that the quality of the results,
objectively measured  in terms of Rand and Jaccard distances between the automatic and the ground truth
segmentations, depends strongly on the capability of the selected metric of capturing the geometrical features of
the mesh. Our experiments with {\it geodesic}, {\it angular} and {\it sdf} distances show that none
is enough to produce good segmentations in all cases. A combination of two o more metrics usually
leads to better results.

Compared to  other segmentation methods considered in the literature, the segmentation method proposed in this paper
has several advantages. First, it does not depend on parameters that must be tuned by hand. Second,
it very is flexible since it can handle any metric to define the distance between triangles.  Finally, it
is very cheap, with a computational cost of $O(nm)$, where $m$ is the cost of computing the distance between
two faces of the triangulation. In this sense, the proposed method is cheaper than spectral segmentation methods, which in the
best case ( when Nystr\"om approximation is used ) compute additionally the eigenvectors of an order $k$ matrix,
with an extra cost of $O((n - k)k^2) + O(k^3)$ operations.

In the present work, we intentionally focus on simple single segmentation fields on $3D$ meshes and the clustering is obtained applying
a well know clustering algorithm, \textit{k-means++}, in order to achieve straightforwardly a fair comparison of our segmentation method with the spectral method. Nevertheless, our farthest point based segmentation {\it FSS} method may be extended to more complicated scenarios, where several attributes are combined in a single segmentation field, such as in \cite{LZSC09}, \cite{WTKL14} or where several multi-view clustering methods  have been proposed to  integrate  without supervision multiple information from the data, \cite{HCC12}, \cite{CNH13}. Therefore, in the future, we plan to investigate the advantages of replacing  Nystr\"om's method with ours, for instance, compare with \cite{YLMJZ12}, \cite{EFKK14}, in order to make the proposed algorithms more efficient in terms of computational and memory complexity as well as to obtain segmentations without artifacts, such as nonconnected clusters.  Finally, it is worth mentioning that the basic ideas of our segmentation method - to compute only few columns of the affinity matrix and to apply a clustering algorithm to the rows of this submatrix - could be straightforwardly used in other segmentation or clustering problems, for instance in segmentation of digital images \cite{MiCh16}.

%

\end{document}